\documentclass[useAMS,usenatbib]{mn2e}

\title{Simple stellar population modeling of low S/N galaxy spectra and quasar host galaxy applications}

\author[G. Mosby, Jr. et al.]{G.~Mosby Jr,$^1$ C.~A.~Tremonti,$^1$ E.~J.~Hooper,$^{1,2}$ M.~J.~Wolf,$^1$ 
\newauthor 
A.~I.~Sheinis,$^3$ and J.~W.~ Richards,$^4$\\
$^1$Astronomy Department, University of Wisconsin-Madison,
    Madison, WI 53706, USA\\
$^2$WIYN Observatory, 950 N. Cherry Ave.,
    Tuscon, AZ 85719, USA\\
$^3$Australian Astronomical Observatory, PO Box 915, North Ryde, NSW 1670, Australia\\
$^4$Mathematics Department, San Francisco State University, San Francisco, CA 94132, USA}

\usepackage{aas_macros}
\usepackage{longtable}
\usepackage{booktabs}
\usepackage[caption=false]{subfig}
\usepackage{graphicx}
\usepackage{amsmath}
\usepackage[countmax]{subfloat}
\usepackage{alltt}
\usepackage{tabularx}
\usepackage{rotating,caption}
\usepackage{natbib}
\usepackage[bottom]{footmisc}
\begin{document}
\label{firstpage}
\maketitle





\begin{abstract}
To study the effect of supermassive black holes (SMBHs) on their host galaxies it is important to study the hosts when the SMBH is near its peak activity. A method to investigate the host galaxies of high luminosity quasars is to obtain optical spectra at positions offset from the nucleus where the relative contribution of the quasar and host are comparable. However, at these extended radii the galaxy surface brightness is often low (20-22 mag per arcsec$^{2}$) and the resulting spectrum might have such low S/N that it hinders analysis with standard stellar population modeling techniques. To address this problem we have developed a method that can recover galaxy star formation histories (SFHs) from rest frame optical spectra with S/N $\sim$ 5~\AA$^{-1}$. This method uses the statistical technique diffusion k-means to tailor the stellar population modeling basis set. Our diffusion k-means minimal basis set, composed of 4 broad age bins, is successful in recovering a range of galaxy SFHs. Additionally, using an analytic prescription for seeing conditions, we are able to simultaneously model scattered quasar light and the SFH of quasar host galaxies (QHGs). We use synthetic data to compare results of our novel method with previous techniques. We also present the modeling results on a previously published QHG and show that galaxy properties recovered from a diffusion k-means basis set are less sensitive to noise added to this quasar host galaxy spectrum. Our new method has a clear advantage in recovering information from QHGs and could also be applied to the analysis of other low S/N galaxy spectra such as those typically obtained for high redshift objects or integral field spectroscopic surveys.
\end{abstract}





\section{Introduction}
A key question in galaxy evolution is: what is the interplay between central supermassive black holes (SMBHs) and their host galaxies? In simulations of galaxy formation, feedback from both an active galactic nucleus (AGN) and star formation accompany the assembly of massive galaxies \citep{Dimatteo-2005, Hopkins-2006, Robertson-2006}. However, different models apply varying prescriptions for these forms of feedback. These prescriptions are often only constrained by reproducing realistic global properties of galaxies such as the observed correlation between the central black hole and host galaxy bulge masses or the galaxy mass function. Thus, to answer questions regarding the effect of SMBHs on their hosts, it is necessary to look for direct observational evidence of feedback.

Rapid quenching of recent star formation is predicted by models of quasar feedback \citep[e.g.,][]{Dimatteo-2005,Hopkins-2008ii}. Observationally, recent quenching can be identified by analysing and fitting galaxy spectra to recover galaxy star formation histories \citep[see][]{Cano-Diaz-2012}. However, looking for direct observational evidence of the effects of feedback from an accreting black hole on its host galaxy is challenging. To establish whether accreting SMBHs have a significant impact on the evolution of their hosts, it is essential to study galaxies when their black holes are near the peak of their activity: the quasar phase. Unfortunately, the scattering of light from the high luminosity quasi-stellar object (QSO) limits the analysis that can be done on the host galaxies. This scattered quasar light is caused by the turbulence of the Earth's atmosphere and the outer wings of the point-spread function of the QSO. A common approach to circumvent this is studying the hosts of low luminosity obscured narrow-line AGN. Recent work, however, suggests that low and high luminosity AGN are different phases of galaxy evolution \citep{Schawinski-2009,Trump-2013}. Work on bona fide quasar host galaxies is mostly limited to observations of the host galaxies outside their centres via offset longslit or integral field unit (IFU) observations, e.g.,  \citet{Nolan-2001,Miller-2003, Jahnke-2004b, Wold-2010,Cano-Diaz-2012}. Because these types of observations probe the low surface brightness outskirts of the host galaxy light profile, spectra obtained are often low to moderate signal-to-noise (S/N $\la$ 20~\AA$^{-1}$).
Scattered quasar light complicates these observations since it must be modeled or subtracted off from the host galaxy observations. In order to recover useful and reliable information from this type of data, a method of spectral fitting must be developed that can handle low signal-to-noise data and model the scattered quasar light.

The challenge of analyzing any galaxy spectrum is to reliably decompose the integrated light from perhaps billions of years of stellar evolution with only a snapshot. A lot of work has been done in this field recently (see \citet{Walcher-2011} for a review). Current methods used for analysis in the rest frame optical include (i) using spectral indices, measured equivalent widths of stellar absorption features \citep{Worthey-1994,Trager-1998,Thomas-2004},  (ii) principle component analysis (PCA) \citep{Murtagh-1987,Connolly-1995,Madgwick-2003, Lu-2006}, and (iii) spectral fitting by inversion, inverting an observed galaxy spectrum onto a basis of independent components  \citep{Heavens-2000,Tremonti-2004,Tojeiro-2007,Ocvirk-2006,Cid-2005,Walcher-2006,Chilingarian-2007,Koleva-2009,Richards-2009}. 

These methods are not optimal for quasar host galaxy studies for the following reasons. Firstly, using spectral indices necessarily requires a high signal to noise (S/N $\ga$ 30~pixel$^{-1}$) spectrum to properly measure equivalent widths and recover galaxy properties \citep{Johansson-2012}, so this method would not be suitable for analyzing low signal-to-noise spectra. Secondly, though PCA--which seeks to decompose a galaxy's spectrum into a linear combination of calculated orthogonal principle components--has been used to analyse low signal-to-noise spectra \citep[e.g.,][]{Chen-2012}, it can be difficult to determine what information about galaxy properties is encoded in the derived principle components. Additionally, within the PCA method, one could not force the scattered QSO light to be a separate PCA component. Thirdly, spectral fitting by inversion assumes a galaxy's spectrum can be decomposed into the sum of the light from single age, single metallicity populations of stars, or simple stellar populations (SSPs). These SSPs represent instantaneous bursts of star formation at different moments in time and their linear combination can represent the star formation history of a galaxy. 
However, most spectral fitting by inversion methods depend on moderate to high signal-to-noise (S/N $\ga$ 20~ pixel$^{-1}$) for reliable results \citep{Mathis-2006,Tojeiro-2007}.\citet{Tojeiro-2007} find that even with \textsc{VESPA} it is difficult to recover meaningful information from individual galaxy spectra with S/N $\la$~10  pixel$^{-1}$ even though \textsc{VESPA} is a spectral analysis program designed to robustly recover star formation histories by adaptively changing the number and binning of SSPs given a spectrum's noise.

We develop and test a new spectral fitting by inversion method for recovering star formation histories from low signal-to-noise galaxy spectra that includes the ability to model scattered quasar light. This new method is set apart from the aforementioned techniques in its use of \textit{diffusion k-means}, a dimension reduction technique that allows us to look at all the spectral information available in a library of SSPs and meaningfully group them \citep{Lafon-2006,Richards-2009}. This property is attractive because it can provide a quantitative way to reduce the number of SSPs used for spectral fitting from a SSP library. It is common to only use a subset of the SSPs (which we will call a basis set) from a library to increase computational efficiency and reduce the use of extraneous SSPs that might have very similar spectra and thus be indistinguishable to any fitting routine. But, most subsets have been chosen empirically or essentially by hand for specific applications \citep{Tremonti-2004,Cid-2005,Tojeiro-2007}. Diffusion k-means provides a quantitative way to form a more manageable basis set from a large SSP library, reducing the degrees of freedom in the fitting and the number of bases to a number $k$ of our choosing. Our method also includes the option to model simultaneously the stellar populations and the quasar scattered light, in which the scattered light is modeled analytically assuming Gaussian seeing during the observations.

Diffusion k-means has already been shown in \citet{Richards-2009} to be effective in forming bases for stellar population modeling of a sample of galaxy spectra from the Sloan Digital Sky Survey I \citep{York-2000}. In this paper, we explore whether using a reduced basis set can improve the accuracy with which galaxy star formation histories are recovered in low S/N data (S/N $\sim$ 5~\AA$^{-1}$).  We compare bases derived using k-means (and some physical intuition) to bases selected using other techniques. We outline a method for modeling scattered quasar light, and we test the accuracy with which stellar population parameters can be recovered in low S/N quasar host galaxy spectra with our k-means basis. We test this by generating synthetic galaxies with a range of given star formation histories and comparing the recovered star formation histories of different basis sets with the input star formation histories. We also use our method on a previously published quasar host galaxy, and compare the results with the literature.

Section 2 describes the spectral fitting technique and introduces the basis sets to be tested. In \S{3}, we describe the star formation histories tested and the results. After testing on galaxy spectra with no scattered quasar light, we then apply our method to some synthetic quasar host galaxy spectra and an observed quasar host galaxy spectrum in \S{4}. We discuss the accuracy and reliability in recovering star formation histories of the tested basis sets in \S{5}. We use the following cosmology $H_{0}=70.0~$ km s$^{-1}$ Mpc$^{-1}$ and a flat universe: $\Omega_{m} = 0.3$, $\Omega_{\Lambda} = 0.7$.

\section{Method} \label{sec:method}
We aim to test if we can improve the recovery of star formation histories in low signal-to-noise spectra by reducing the number of bases in spectral fitting. We generate synthetic galaxy data with 6 star formation histories (details in \S{\ref{testedsfhs}}). Next, we fit this data using 4 basis sets (detailed in \S\ref{testedsets}) and compare the recovered star formation histories to the input star formation histories to examine the relative effectiveness of each basis set. To recover the star formation histories from galaxy spectra, we use a spectral fitting by inversion technique described in \S\ref{specfit}.

\subsection{Spectral Fitting and Implementation}\label{specfit}
To fit and recover a star formation history from a given observation of a galaxy, we use a modified code originally written by \citet{Sheinis-2002} and later developed by \citet{Wold-2010} that will be henceforth referenced as \textsc{sspmodel}. The inputs for the code are spectroscopic observations of a galaxy, and optionally, a second spectroscopic observation of a central QSO. The user also provides a set of bases to use in fitting the galaxy spectrum. \textsc{sspmodel} then finds the best fit to the input galaxy spectrum by performing a weighted least-squares minimization ($\chi^2$ minimization) using alternating simulated annealing and downhill simplex minimization routines outlined in \citet{NR}. The output of the code is an estimate of the relative light fraction of each stellar population basis, the galaxy's \textit{V}-band attenuation, $A_V$ using the extinction curve of \citet{Cardelli-1989}, and if requested, the parameters that best describe any scattered quasar light. For this paper, the code was rewritten in C for speed and portability. Previous incarnations of the code in Interactive Data Language (IDL) ran 5 times slower and required a license to run. We have also updated the treatment of scattered quasar light.  The formalism and equations describing the model fitting and extinction are detailed in \citet{Wold-2010}. The details and method of estimating the scattered quasar light are in \S\ref{scattqso}.

In our code, \textsc{sspmodel}, we make the general assumption of the spectral fitting by inversion technique that a galaxy spectrum can be decomposed into linear combinations of stellar populations that represent the galaxy's star formation history. A common technique to estimate these stellar populations is to assume the galaxy spectrum can be modeled by the linear combination of a set of SSPs and some dust attenuation. These SSPs are the spectra of a single age, single metallicity population of stars at given times after an instantaneous burst of star formation. These SSPs are normalized to have formed a fixed amount of stars initially so that by decomposing a galaxy into weighted combinations of them, we can estimate how much star formation occurred at a given time--the star formation history. If we had an infinite number of SSPs so as to sample time as finely as possible, we could theoretically recover very detailed star formation histories. But, since the youngest stars typically dominate the light in an integrated spectrum for galaxies with active star formation, older SSPs would contribute very little to the integrated spectrum and our ability to get precise star formation histories at large lookback time would be hindered. Additionally, the SSP spectra change very little several Gyr after the initial burst, so including more older SSPs might introduce degeneracies in the fitting. Thus, more SSPs to cover more time steps is not a panacea. In practice, one chooses a subset of SSPs to fit a galaxy spectrum.

For all of the analysis of galaxy spectra in this paper we use SSPs from \citet{Bruzual-2003} (henceforth referred to as BC03). Specifically, we use the high resolution, solar metallicity Padova 1994 instantaneous burst models with an initial mass function (IMF) from \citet{Chabrier-2003}. These SSPs rely on the STELIB stellar library \citep{Leborgne-2003}. The STELIB library has a wavelength coverage of 3200 -- 9500~\AA$ $ at a spectral resolution of $\la$3~\AA$ $ sampled at 1~\AA. The SSPs in BC03 have a large wavelength coverage, but for all the analysis in this paper we restrict ourselves to 3600 -- 8500~\AA$ $ since this is the approximate range of our quasar host galaxy observations. BC03 includes SSPs for 6 metallicities, but we limit our tests to solar metallicity as a starting point. The STELIB library also has the most template stars for solar metallicity, making the BC03 derived spectra at this metallicity more accurate. Additionally, the young stellar populations in galaxies are less sensitive to the age metallicty degeneracy so that even though we expect galaxies to host a range of populations at different metallicities, the mono-metallicty assumption suffices for galaxies with recent star formation. \citet{Brotherton-1999} showed for a post starburst quasar (PSQ), the uncertainties caused by assuming a single metallicity were $\pm$ 50 Myr on a 400 Myr population. Similar assumptions have also been made in recent PSQ studies \citep[e.g.][]{Cales-2013}.

Under our spectral fitting by inversion assumptions that we can fit a galaxy's spectrum as the linear sum of solar metallicity SSPs, we can think of the luminosity of a galaxy at a given wavelength being determined by:

\begin{equation}
\label{eq:sspsum}
L_{\mathrm{gal}}(\lambda) = 10^{-0.4 A_{\lambda}}\sum_{i} a_{i} S(\lambda,Z_{\odot},t_i)
\end{equation}
where $L_{\mathrm{gal}}(\lambda)$ is the galaxy luminosity at wavelength $\lambda$, $ S(\lambda,Z_{\odot},t_i)$ is the luminosity of a solar metallicity, age $t_i$ SSP at wavelength $\lambda$, and the $a_i$'s are the coefficients representing the amount of stellar mass formed $t_i$ years ago. The extinction at each wavelength, $A_{\lambda}$, is calculated using the reddening law of \citet{Cardelli-1989}, and is assumed for simplicity to be the same for all stellar populations with $R_{V}$ = 3.1. In BC03, the SSPs are normalized to 1 M$_{\odot}$, so the $a_i$'s represent the number of solar masses formed $t_i$ years ago.\footnote{In this paper, we are primarily interested in the mass formed (i.e. the star formation history). However, the present day stellar mass can be trivially calculated by using constants tabulated in BC03 that give fraction of the mass formed that is still in stars for a SSP at a given age.}

Framing the spectral fitting by inversion problem as it is in Eqn. \ref{eq:sspsum} assumes that  $a_{i}$ M$_{\odot}$ of stars were formed instantaneously at a time $t_i$ ago for a galaxy. This star formation history is the sum of delta functions. There is a natural uncertainty in this method--what if a galaxy is composed of stars whose ages are not represented by the SSPs chosen? Any minimization technique would simply seek the best combination of the given SSPs to reproduce the galaxy spectrum, but it is unpredictable without testing how the minimization technique will account for unrepresented populations using an incomplete set of bases. We can eliminate the need to cherry pick only a handful of individual SSPs by using diffusion k-means. We use diffusion k-means to identify similar SSPs, group them into $k$ groups, and then we perform a weighted average of the SSPs in each group to form a new reduced basis set of $k$ bases. 

\subsection{Tested Basis Sets} \label{testedsets}
To test whether lowering the number of bases used to model the stellar populations present in a galaxy spectrum allows us to reliably and accurately recover the star formation history with low signal-to-noise data, we run synthetic galaxy spectra through \textsc{sspmodel} using 4 distinct basis sets (Table \ref{tab:test-bases-ages}): the 15 solar metallicity SSPs used by \citet{Cid-2005} (hereafter referred to as the `traditional' basis set, or TB), a diffusion k-means selected set (DFK-AVG), a constant light fraction set (CONSTLF), and a set defined by the individual SSPs with ages closest to the average age of the DFK-AVG bases (DFK-SSP). Each basis set is described in detail below.

\begin{figure*}
\begin{center}
\includegraphics{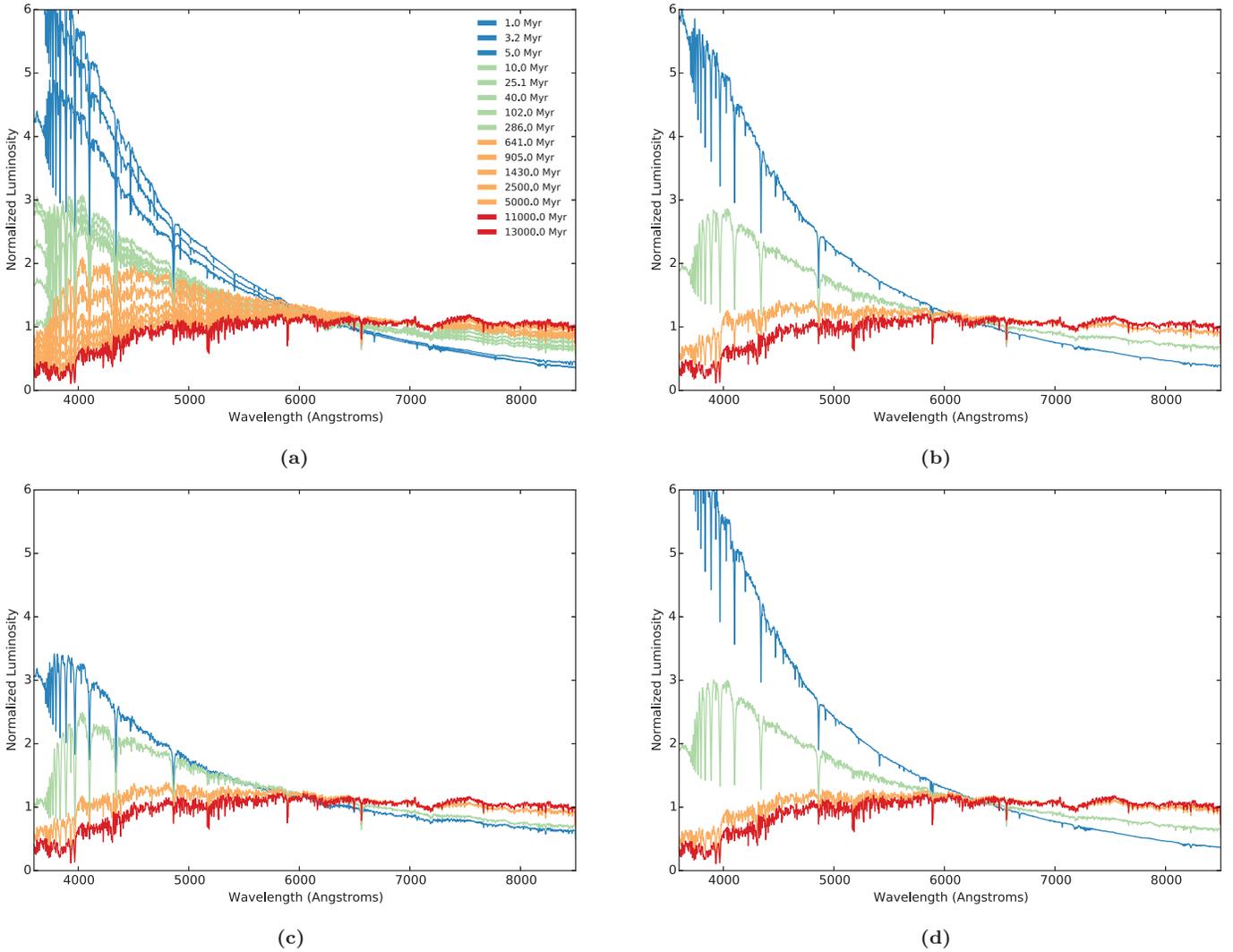}
\end{center}
\caption{\label{fig:multi_panel_basis} The model spectra of the four basis sets tested. Each spectrum is normalized by its median luminosity. (a) The traditional basis set (\S\ref{tb}). Solar metallicity SSPs from \citet{Cid-2005}. The ages of each SSP are shown in the legend. (b) The diffusion k-means basis set (\S\ref{diffusionkmeans}). (c) The constant light fraction basis set (\S\ref{constlf}). (d) The DFK-SSP basis set composed of SSPs closest to the average age of the DFK-AVG groups (\S\ref{midbin}). The spectra are color coded by the diffusion k-means group they belong with group 1 in blue, group 2 in green, group 3 in orange, and group 4 in red.}
\end{figure*}

\subsubsection{Traditional Basis Set} \label{tb}
Traditionally in the spectral fitting by inversion method, one chooses individual SSPs from a large library of SSPs for spectral fitting and recovering star formation histories. The SSPs are often chosen empirically and might be tailored to specific goals \citep{Tremonti-2004,Tojeiro-2007,Cid-2005}. The 15 solar metallicty SSP ages chosen by \citet{Cid-2005} have been used by several groups \citep[e.g.,][]{Wold-2010,Richards-2009} to fit galaxy spectra. We call these 15 SSPs our traditional basis set (TB). This set serves the purpose of a control in our tests. By comparing the results of this TB basis set and the smaller basis sets, we test whether a smaller basis set can recover a range of star formation histories as well as this larger, more traditional basis set. To compare the recovered star formation histories, we bin the TB derived masses into four age bins chosen by diffusion k-means discussed in \S \ref{diffusionkmeans}. Fig. \ref{fig:multi_panel_basis}a shows the traditional basis set's model spectra and their corresponding ages.

\begin{table*}\centering
\caption{Table of the age ranges spanned by the basis sets tested and light fractions of the diffusion k-means (DFK-AVG) and constant light fraction (CONSTLF) basis sets. The ages of the SSPs selected for the DFK-SSP basis set are shown in the last column. The age bin abbreviations and diffusion k-means groups referred to in the text are labelled. The light fractions are the fraction of light a base spectrum contributes summed over 3600-8500 \AA$ $ relative to the the total light contributed by a basis set's spectra as demonstrated in Eqn. \ref{cfrac}.}
\noindent\makebox[\textwidth]{
\begin{tabular}{@{}cccccccccc@{}}\toprule
\multicolumn{2}{c}{DFK Group}&\multicolumn{2}{c}{Age Abbrev.}&\multicolumn{2}{c}{DFK-AVG} & \multicolumn{2}{c}{CONSTLF} & \multicolumn{2}{c}{DFK-SSP} \\
\midrule
\multicolumn{2}{c}{}&\multicolumn{2}{c}{}&Ages & Light Fraction & Ages & Light Fraction&Ages\\
\multicolumn{2}{c}{1}&\multicolumn{2}{c}{Y}&0.9 -- 5.2 Myr &0.06& 0.9 -- 101.5 Myr &0.25& 2.5 Myr \\
\multicolumn{2}{c}{2}&\multicolumn{2}{c}{I1}&5.5 -- 404 Myr&0.36 & 113.9 -- 718.7 Myr&0.25&  64.1 Myr\\
\multicolumn{2}{c}{3}&\multicolumn{2}{c}{I2}&453.5 -- 5750 Myr&0.43& 806.4 -- 3250 Myr&0.25& 2500 Myr\\
\multicolumn{2}{c}{4}&\multicolumn{2}{c}{O}&6000 -- 13500 Myr&0.16& 3500 -- 13500 Myr&0.26& 9750 Myr\\
\bottomrule
\end{tabular}
}

\label{tab:test-bases-ages}
\end{table*}

\subsubsection{Diffusion K-Means Basis Set Formation} \label{diffusionkmeans}
The power of diffusion k-means is that one can run any multidimensional data set (in our case a set of SSP spectra where each wavelength is a different dimension) through the algorithm and receive low dimensional coordinates for each multidimensional point (an SSP spectrum) that encode how similar the multidimensional points in the data set are. This process is called diffusion mapping. We call the lower dimension space the mapped points occupy the diffusion space. Diffusion mapping compares each multidimensional point to all the other points in the data set. Thus, diffusion k-means is sensitive to detecting amplitude and shape differences between multidimensional points. As a consequence, depending on the goals of running diffusion k-means, one might need to normalize or pre-process the multidimensional data set to remove fixed scale offsets that would show up as differences between the data (spectra) before running the algorithm. After diffusion mapping the multidimensional points in a data set, one can then perform the k-means clustering algorithm in the diffusion space to group the most similar data points into $k$ groups. 

Our first step in creating a diffusion k-means basis set for spectral fitting is to select which SSPs will be mapped then grouped. We found that the age 0.125 Myr - 0.891~Myr  SSPs in the Chabrier IMF, solar metallicity subset of BC03 have identical spectra and did not include SSPs younger than 0.891~Myr, leaving 203 solar metallicity SSPs from BC03. We also perform a time cut, restricting ourselves to SSPs with ages younger than or equivalent to the current age of the universe (13.5~Gyr) leaving 177 SSPs. 

Since we want to group SSPs based on stellar features and spectral shape not amplitude (naturally the youngest SSPs have the highest luminosities), our second step is to normalize each SSP by its average luminosity in the wavelength range 3600 -- 8500~\AA$ $ before running them through the diffusion k-means algorithm. We found that there were differences in the final diffusion k-means groupings depending on whether we normalize by the median or average, and we settled on using the average since the bases formed using the median normalization did not seem to represent a variety of spectral shapes as well as the bases formed using the average normalization. Because diffusion k-means is sensitive to normalization and wavelength coverage, interested readers should explore what works best for their own applications and tailor diffusion k-means appropriately.

Our selected and normalized SSPs are next used to form a diffusion map. We use the \verb+diffuse+ function from the \verb+diffusionMap+ package, written in \textsc{R} by Joseph Richards \citep{R,Richards-2009}. This function takes an additional tuning parameter in the mapping, $\epsilon$, the diffusion constant. The parameter $\epsilon$ modifies the shape of the distribution of points (SSPs) in the diffusion space. We choose this constant empirically to qualitatively minimize outliers, smooth the distribution of points in the diffusion space, and ensure that the ages of the constituent SSPs do not overlap from one group to another. We found empirically that $\epsilon = 70$ worked well by generating several diffusion maps and only varying epsilon until the desired conditions above were met.  For a different multidimensional data set, one would have to tune $\epsilon$ again.

\begin{figure*}
\begin{center}
\includegraphics{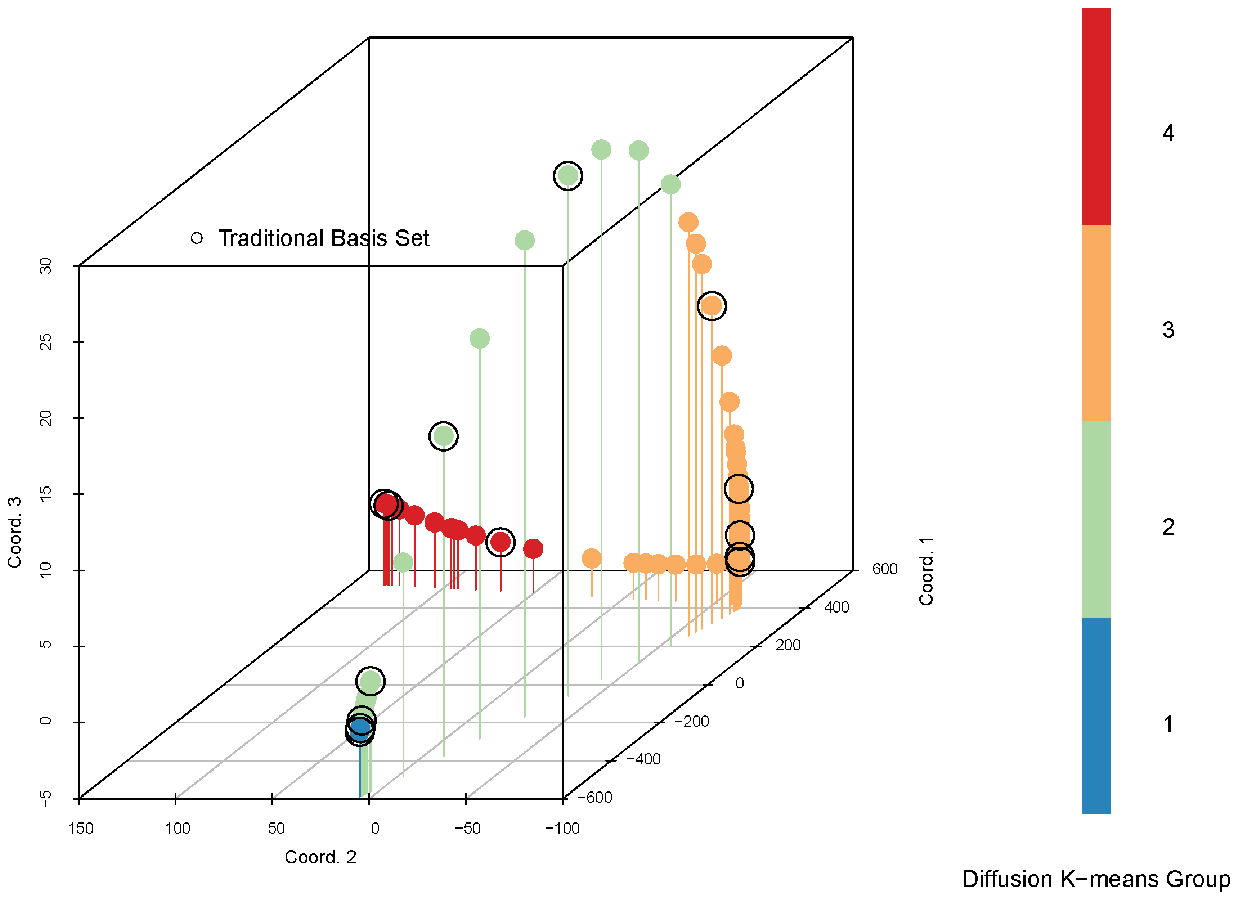}
\end{center}
\caption{The 177 solar metallicity Chabrier IMF BC03 spectra in the derived diffusion space. The points are color coded by k-means group, which is an age sequence as seen in Table \ref{tab:test-bases-ages}. The 15 SSPs used by \citet{Cid-2005}, our `traditional' basis set, are marked by black circles. The traditional basis set selects SSPs that diffusion k-means finds are very similar, especially for groups 2 and 3. This can result in more degenerate spectral fitting. Also note that the SSPs in the youngest age bin (group 1) all have very similar diffusion map coordinates, and thus appear very clustered.}
\label{fig:dfkmap}
\end{figure*}

Our final diffusion map reduces the 4901 (number of wavelengths) dimension data to 3 dimensional coordinates for each SSP. In this 3-D space, the closer two points (SSPs) lie to each other the more similar they are. The 177 SSPs are plotted in this diffusion space, color coded by their final k-means group in Fig. \ref{fig:dfkmap}. Additionally, the individual SSPs used by \citet{Cid-2005} are circled. The first k-means group has the youngest average age. The fourth k-means group has the oldest age.

Using the diffusion map of the 177 SSPs, we can then group similar SSPs together. We use the \verb+diffusionkmeans+ function from the \verb+diffusionMap+ package, which uses k-means, to group the SSPs into $k$ groups inside the diffusion space. Each of these groups is used to form a new base spectrum in the basis set for modeling stellar populations by averaging the individual SSPs in the group as described below.

Our interest in modeling the stellar populations and scattered quasar light of low signal-to-noise spectra drive us to choose as small a $k$ as possible while remaining physically interesting and capable of recovering an accurate star formation history. We initially tried $k=3$, but with $k=3$, the diffusion k-means basis set had trouble fitting strongly peaked star formation histories and recovering the correct reddening of the host galaxy, even for high signal-to-noise data. Consequently, we increased $k$ to 4.

\begin{figure}
\centering
\includegraphics{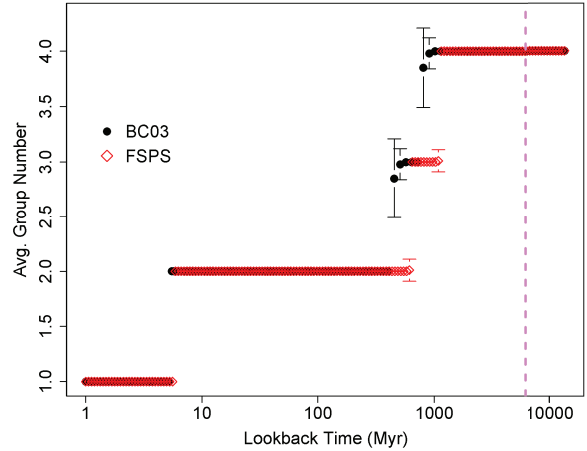}
\caption{The average group assignment (1st -- 4th) for SSPs included in the diffusion k-means process. The black circles are the BC03 SSPs analyzed in this paper. The error bars show the standard deviation of an SSP's group assignment. The dashed vertical line shows 6~Gyr. For reasons given in \S\ref{diffusionkmeans} we decided to allocate SSPs younger than 6~Gyr to the 3rd group and SSPs at 6~Gyr and older to the 4th group. The diffusion k-means groupings will necessarily depend on the input SSP spectra, and there are several SSP models available. For a comparison, the red diamonds show the solar metallicity, Chabrier IMF SSPs from the Flexible Stellar Population Synthesis code \citep{Conroy-2009,Conroy-2010} and how diffusion k-means groups them.}
\label{fig:grpstab}
\end{figure}

It is worth noting that diffusion k-means does not necessarily return the same groupings of the SSPs in multiple runs for a given $k$ because of the nature of the k-means algorithm. In the k-means step, $k$ centroids are randomly chosen in the diffusion map. Each SSP is then assigned to belong to the group of the closest centroid. For the $k$ groups of SSPs, updated centroids are then calculated. This process of assigning and updating is repeated for 10 iterations. Because of the stochasticity of the initialization, it is possible to get different groupings. However, if the k-means algorithm has converged on a local optimum, the k-means groupings should not change. In Fig. \ref{fig:grpstab}, we show the mean diffusion k-means group assignments of the included SSPs for 100 runs of the  \verb+diffusionkmeans+ function for $k = 4$. The standard deviations of the group assignments are plotted as vertical error bars. Fig. \ref{fig:grpstab} shows that the assignments of the first two groups are very stable. In fact, the least stable assignments only happen on the endpoints of the third group, but these assignments only affect the grouping of 4 SSPs out of 177 total. 

In this work, we use the BC03 solar metallicity SSPs, but in principle the diffusion k-means groupings might change for a different set of SSPs. To explore this, we plot in Fig. \ref{fig:grpstab} the solar metallicity, Chabrier IMF SSP models produced by the Flexible Stellar Population Synthesis (FSPS) code \citep{Conroy-2009,Conroy-2010}. We chose the FSPS models as they include ages as young as the youngest SSPs from BC03 and have a similar spectral resolution. Though BC03 uses the STELIB stellar library with Padova 1994 tracks and FSPS uses the MILES stellar library \citep{Sanchez-Blazquez-2006} with Padova 2000 tracks, we find similar results. This suggests that at least for the wavelength and metallicity we've chosen, the diffusion k-means groupings are more sensitive to the differences between stellar populations than differences between the SSP models.

Once the 177 SSPs are grouped by k-means into 4  groups, we perform a weighted average of all the SSPs in each group. Performing this weighted average normalizes each new diffusion k-means base to form 1 M$_{\odot}$ in the time spanned by the SSPs composing the average. Revisiting the formalism from \S\ref{specfit}, in the case of using the reduced basis set, we can think of the luminosity of a galaxy at a given wavelength by:

\begin{equation}
\label{eq:basesum}
L_{\mathrm{gal}}(\lambda) = 10^{-0.4 A_{\lambda}}\sum_{k'} a_{k'} D(\lambda,Z_{\odot},\bar{t}_{k'})
\end{equation}

\begin{equation}
\label{eq:dfkavg}
D(\lambda,Z_{\odot},\bar{t}_{k'}) = \frac{\sum\limits_{i \in k'} w_i S(\lambda,Z_{\odot},t_i)}{\sum\limits_{i \in k'} w_i}
\end{equation}
where $D(\lambda,Z_{\odot},\bar{t}_{k'})$ is the luminosity of a weighted average of the solar metallicity SSPs in the $k'$th diffusion k-means group at wavelength $\lambda$ (Eqn. \ref{eq:dfkavg}). The $a_{k'}$'s still represent the number of solar masses formed but now over the time spanned by the $k'$th base. Note that though we normalize the SSPs by their average luminosity for diffusion mapping, the SSPs are not normalized to form each of the $k$ basis spectra.

In the new formalism of Eqn. \ref{eq:dfkavg} with averaged SSPs, the weights, $w_i$, in the averaging are defined to be the time spanned between the two nearest \textit{midpoints} of adjacent SSPs, given in Eq. \ref{delta-t} with $t$ corresponding to the age of the SSP. The endpoints are treated differently. The first SSP's weight is defined as the time between the present and the midpoint between the first 2 SSPs. The last SSP's weight is defined as the time between the midpoint between the last 2 SSPs and the age of the universe, which is the age of the last SSP.
\begin{equation}
\label{delta-t}
w_{i} = \left\{
	\begin{array}{ll}
		\frac{t_{1}+t_{2}}{2}, \mbox{ for } i = 1 \\[6pt]
	           \frac{t_{i+1}-t_{i-1}}{2}, \mbox{ for } i \neq 1 \mbox{ and }  i \neq N \\[6pt]
		\frac{t_{N}-t_{N-1}}{2}, \mbox{ for } i = N
	\end{array}
	\right.
\end{equation}
The age of the first SSP corresponds to $t_{1}$, and $N$ is the total number of SSPs. $t_{N}$ is 13.5~Gyr. How the SSPs are combined is a choice we must make. We use a weighted average with the weights in Eqn. \ref{delta-t} as opposed to a simple arithmetic mean because the SSPs in BC03 are not evenly spaced in time. Using the weights defined in  Eq. \ref{delta-t}, our average of the SSPs in a k-means group approximates the integrated spectrum of stars formed between the oldest and youngest SSP of each k-means group assuming star formation has been constant across the time interval spanned by the base. The 4 bases formed from this weighted average are assumed to have the average properties of the SSPs comprising them.

Looking at the diffusion k-means groupings in Fig. \ref{fig:grpstab}, we noticed that the oldest group spanned $\sim$10~Gyr. Though it is reasonable that diffusion k-means would group all of the age $\ga$1.0~Gyr SSPs together (their spectra are very similar), it has been shown that star formation histories are not usually constant over such long time spans \citep{Pacifici-2013}. As a quick remedy to this, we simply divide the last group into 2 groups at its approximate midpoint, 6~Gyr (a dashed pink line in Fig. \ref{fig:grpstab} marks this time). SSPs younger than 6~Gyr move into the 3rd group. SSPs older or equal to 6~Gyr in age move to the last group. The groups of spectra are then averaged as mentioned above. These 4 bases are then used as the diffusion k-means basis set hereafter referred to as the DFK-AVG basis set. Table \ref{tab:test-bases-ages} lists the age ranges of the bases formed from diffusion k-means alongside the other basis sets to be tested. This set of model spectra is shown in Fig. \ref{fig:multi_panel_basis}b.

Diffusion k-means provides a quantitative description of how similar our initial set of 177 SSPs are to one another. This gives us the freedom to meaningfully reduce the basis set size to $k$=4 for our anticipated application of analyzing low signal-to-noise data. Because we can now confidently group similar SSPs, by performing a weighted average of them, we avoid choosing a subset of individual SSPs. This choice, however, is traded for another. By using weighted averages of the individual SSPs, we are assuming that the star formation history is constant over the time spanned by the SSPs in each base. Consequently, instead of modeling a galaxy with a star formation history composed of delta functions, we are now modeling a galaxy with a star formation history that is continuous, but constrained to be constant from $t$= 0 -- 5~Myr, 5 -- 404~Myr,  0.4 -- 5.7~Gyr, and 6 -- 13.5~Gyr. It would be possible to adjust the weighting on the diffusion k-means bases to represent any given SFH, but we choose constant as a reasonable ``maximum ignorance" starting point. For the first and second bases, the time scales spanned ( $\Delta t$ = 5 and 400~Myr) are short enough for constant SFR to be a reasonable approximation for most galaxies. For the third and fourth bases, $\Delta t$ is large ($\sim$6~Gyr) so our assumption of a constant SFH may be unrealistic. However, since the SSP spectra evolve very gradually at these ages, the weighted average basis is only weakly sensitive to the assumed SFH. In \S{\ref{basistesting}} we will test how well our diffusion k-means bases can represent realistic galaxy SFHs. Diffusion k-means offers useful quantitative constraints for spectral fitting basis set selection while still allowing some liberties to the user.

\subsubsection{Constant Light Fraction Basis Set Formation} \label{constlf}
In the case of modeling a galaxy spectrum with a limited number of bases, one concern might be that one spectrum among the chosen set of bases might not contribute a significant fraction of light in a typical galaxy spectrum. Such a base would be extraneous. To test whether analyzing data with a basis set whose constituents contribute equal light yields significantly different results than the DFK-AVG basis set, we form the constant light fraction (CONSTLF) basis set. We specifically choose this basis set to have equal light fractions for each base's spectrum. The light fraction of an SSP or base is calculated by taking the ratio of the total integrated light of the spectrum (Eqn. \ref{norm}) in question and the sum of the integrated light of all the other SSPs or bases according to Eq. \ref{light-frac} over the wavelength range 3600 -- 8500~\AA.

$N_i$ is defined as the integrated luminosity for wavelengths between and including 3600 -- 8500~\AA$ $ for the $i$th spectrum:
\begin{equation}\label{norm}
N_i = \sum \limits_{\lambda} S(\lambda,Z_{\odot},t_i)
\end{equation}

The light fraction is the following with the $w_i$'s as the weights defined in Eqn. \ref{delta-t}.
\begin{equation} \label{light-frac}
f_{i} = \frac{ N_{i} w_i}{ \sum \limits_{i} N_{i} w_{i}}
\end{equation}

We group the 177 SSPs sequentially by age into 4 groups that each contribute approximately one quarter of the light using the same time averaging described for the DFK-AVG basis set. For the $k'$th CONSTLF basis set spectrum, $C(\lambda,Z_{\odot},\bar{t}_{k'})$, the light fraction is as follows:
\begin{equation} \label{cfrac}
\frac{ \sum \limits_{i \in k'} N_{i} w_i}{ \sum \limits_{k'}\sum \limits_{i \in k'} N_{i} w_{i}} \approx 0.25
\end{equation}
The derived age ranges for the CONSTLF basis set along with the similarly derived DFK-AVG light fractions are listed in Table \ref{tab:test-bases-ages}. The CONSTLF basis set of model spectra is shown in Fig. \ref{fig:multi_panel_basis}c. Note that the CONSTLF basis set's youngest spectrum has a different shape than the other 3 basis sets. This is because the CONSTLF basis set averages together more young SSPs over a larger age bin.

\subsubsection{DFK-SSP Basis Set Formation} \label{midbin}
For the DFK-AVG and CONSTLF basis sets, we average groups of SSPs. But, to test if using a low number of single SSPs would be sufficient, we also form a basis set using the individual SSPs closest to the mean ages of the 4 DFK-AVG bases. The ages of the 4 single SSPs selected in this manner are listed in Table \ref{tab:test-bases-ages}. This DFK-SSP set of model spectra is shown in Fig. \ref{fig:multi_panel_basis}d and are hereafter referred to as the DFK-SSP basis set. 

\begin{figure*}
\begin{center}
\includegraphics{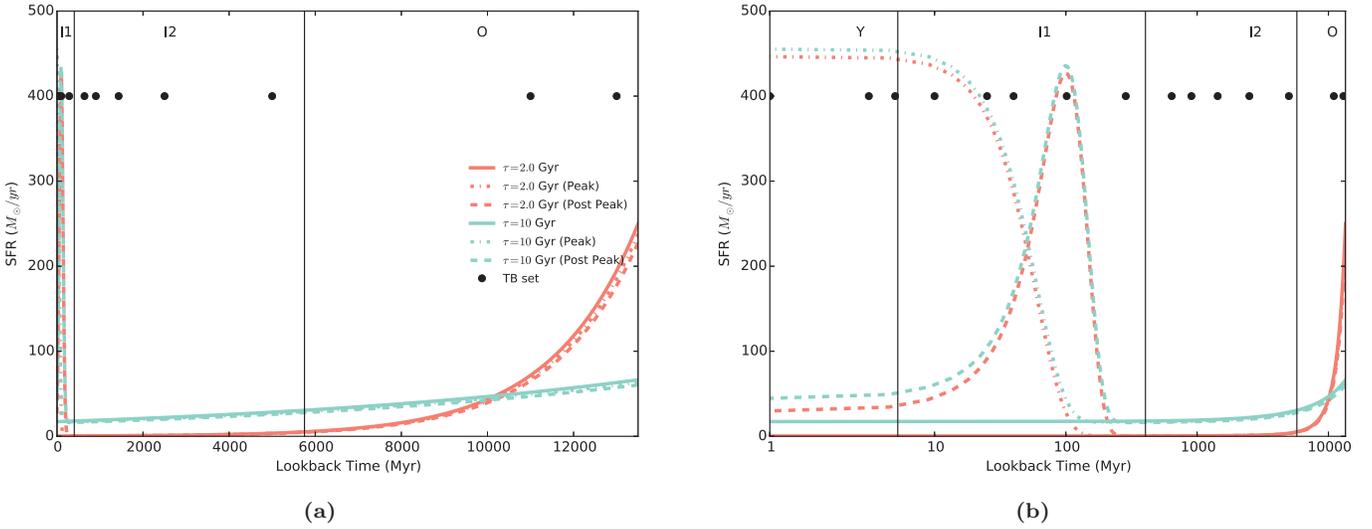}
\end{center}
\caption{The star formation histories (SFHs) used in the fiducial test cases normalized to have formed 5 $\times$~10$^{11}$ M$_{\odot}$ by the present. The solid lines represent a strongly peaked SFH ($\tau=$2~Gyr) and an almost constant SFH ($\tau=$10~Gyr) which are two limiting cases. The dot-dashed lines represent the $\tau$-model star formation histories with a Gaussian burst (10 per cent of total stellar mass formed) whose peak is occurring now. The dashed lines represent the $\tau$-model SFHs with a Gaussian burst (10 per cent of total stellar mass formed) whose peak occurred 100 Myr ago.The age bins (Y, I1, I2, O) derived from the age ranges of the diffusion k-means basis set are shown by the vertical lines. The ages of the SSPs of the TB basis set are denoted by black dots. These two panels are identical except the horizontal axis is log-scaled in (b). }
\label{fig:sfhs}
\end{figure*}

\section{Basis Testing}\label{basistesting}
Our aim in this paper is to test if we can reliably and accurately recover star formation histories from low signal-to-noise data by decreasing the number of and judiciously selecting bases used in a spectral fitting by inversion technique. Accomplishing this will allow us to comfortably analyse low signal-to-noise data such as that obtained observing quasar host galaxies. We first compare and test the reduced basis sets (described in \S\ref{sec:method}, shown in Fig. \ref{fig:multi_panel_basis}) in the absence of the complexity of scattered quasar light using synthetic galaxy spectra as inputs to \textsc{sspmodel}.

\subsection{	Model Galaxy Star Formation Histories}\label{testedsfhs}
For these basis set comparison tests, we use two exponential star formation histories (commonly called $\tau$ models). These star formation histories assume that the star formation rate for a galaxy decreases exponentially with time with some characteristic e-folding time, $\tau$:

\begin{equation} \label{tau}
\psi(t) \propto \mathrm{e}^{-t/\tau} \mbox{ M$_{\odot}$ yr$^{-1}$}
\end{equation}
where $\psi(t)$ is the star formation rate and $t$ is time.

We select a $\tau = 2$~Gyr model as one of our fiducial star formation histories to represent an early-type galaxy in accordance with the findings of \citet{Thomas-2005}. A galaxy with $\tau=2$~Gyr has a strongly peaked star formation history forming 50  per cent of its stars between 12.1 -- 13.5~Gyr ago and 90  per cent of its stars between 8.9 -- 13.5~Gyr ago. We select a $\tau = 10$~Gyr model as our second fiducial star formation history to represent a late-type galaxy. A galaxy with $\tau = 10$~Gyr has a star formation history that declines more gradually with time. A $\tau=10$~Gyr galaxy forms 50  per cent of its stars between 8.9 -- 13.5~Gyr ago, 90 per cent of its stars between 2.5 -- 13.5~Gyr ago. Both star formation histories are shown in Fig. \ref{fig:sfhs}. The right panel shows the star formation histories with a logarithmic scale for the time axis. The left panel shows the time on a linear scale. Lines mark the edges of the age bins formed using the age ranges of the DFK-AVG basis set listed in Table \ref{tab:test-bases-ages}. 

In addition to testing each basis set's recovery of the two fiducial star formation histories, we also test how well each basis set can recover the star formation history of a galaxy with a recent burst of star formation. Current theories of galaxy evolution predict that star formation accompanies black hole growth \citep[e.g.,][]{Hopkins-2006}, so we might expect to have recent bursts of star formation in quasar host galaxies similar to what has been observed locally \citep[e.g.,][]{Cid-2004,Storchi-Bergmann-2005,Davies-2007}. Moreover, if quasar feedback quickly quenches star formation, this burst will be short. In order to test how well the basis sets recover star formation histories with bursts, we add a Gaussian burst with a full-width-half-max of 100~Myr and a mass fraction of 10 per cent of the total stellar mass formed by the present day to the two fiducial star formation histories. We test two burst star formation histories: one in which the galaxy is observed at the peak of the burst and one in which the galaxy is being observed 100~Myr after the peak of the burst. Note, for the star formation history in which the galaxy is observed at the peak of the burst, the burst mass fraction is 5 per cent of the total mass formed; the remaining 5 per cent would be formed in the future. These star formation histories are also shown in Fig. \ref{fig:sfhs}.

We use BC03's GALAXEV program to generate the integrated spectra of galaxies with the 6 chosen star formation histories at the current age of the universe (13.5 Gyr using our adopted cosmology). We then redden the galaxy spectra according to \citet{Cardelli-1989} with a \textit{V}-band attenuation $A_{V} = 1.0$ for all tests. We then add noise to make the S/N $\sim$ 5~\AA$^{-1}$  at all wavelengths. Three hundred independent noise realizations are generated for each star formation history and basis set. We then run \textsc{sspmodel} on each realization using the Center for High Throughput Computing at the University of Wisconsin-Madison.

To judge the accuracy and precision of recovered quantities for each basis set, we look at histograms (of the 300 realizations) of the reduced chi-square of the fits to the input spectra and 3 measured quantities: (1) the mass-to-light ratio, (2) the \textit{V}-band attenuation, $A_{V}$, and (3) the mass fraction in 4 broad age bins (Y-young, I1-intermediate 1, I2-intermediate 2, and O-old) listed in Table \ref{tab:test-bases-ages}. For the DFK-AVG, CONSTLF, and DFK-SSP basis sets, the mass fractions are simply the ratios of the derived masses for each base to the total stellar mass formed. For the TB basis set, the mass fraction is the ratio of the sum of the recovered mass of each TB base within a given broad age bin to the sum of all the recovered mass by the TB basis set. The results for the $\tau = 2$~Gyr tests are presented in \S\ref{tau2}. Those for $\tau = 10$~Gyr are in \S\ref{tau10}. 

The results for tests using the CONSTLF basis set are discussed separately from the other basis sets in \S\ref{constlf-res}. This is because the mass fractions for the CONSTLF basis set are not divisible into the same broad age bins as the DFK-AVG, TB, and DFK-SSP basis sets. The burst star formation history test results are summarized in \S\ref{tauburst}.

\subsection{Tau = 2 Gyr Model Results}\label{tau2}
\begin{figure*}
\includegraphics{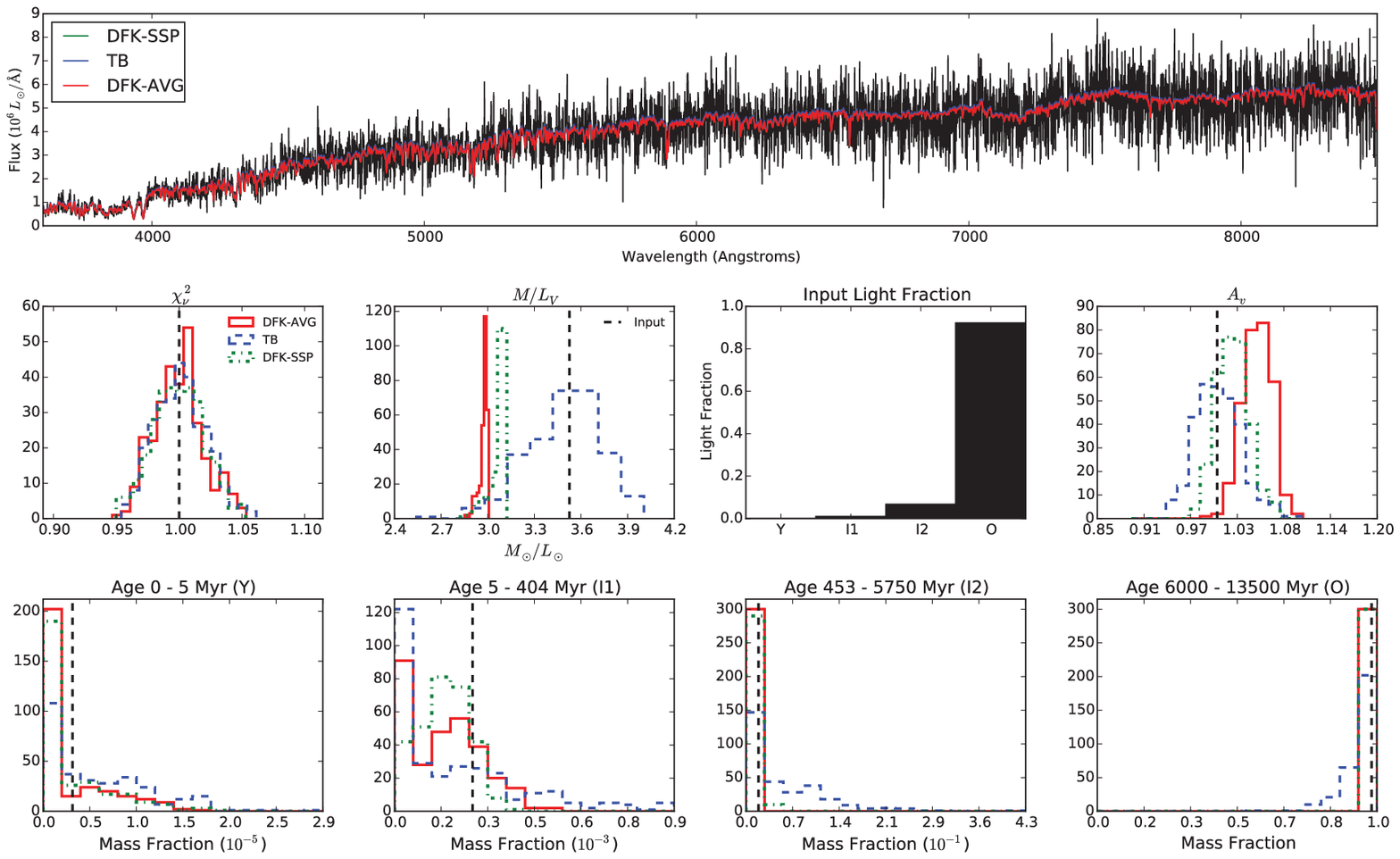}
\caption{The results of running \textsc{sspmodel} on 300 realizations of a synthetic early-type ($\tau = 2$ Gyr) galaxy spectrum with S/N $\sim$ 5~\AA$^{-1}$ for the DFK-AVG, TB, and DFK-SSP basis sets. The top panel shows one of the input spectra in black with the best-fitting spectra for each of the three basis sets overplotted in color.  The three best-fitting spectra are difficult to distinguish by eye as they are quite similar to each other and largely overlap on the plot.  In the middle and lower panels the results of the DFK-AVG basis set are shown as a red solid line; TB as a blue dashed line; and DFK-SSP as a green dot-dashed line.  The middle panel shows histograms of the reduced $\chi^2$, the \textit{V}-band mass-to-light ratio, and the \textit{V}-band dust attenuation.  For reference, we also show the fraction of light in the input spectrum (summed over $\lambda=3600-8500$~\AA) in our four broad age bins: Y~(0--5~Myr); I1~(5--404~Myr); I2~(0.4--5.7~Gyr); and O~(6--13.5~Gyr) from Table \ref{tab:test-bases-ages}.  The bottom row shows the histograms of the recovered mass fractions in each age bin. Note that in all histograms the horizontal axis was chosen to display all values selected by each basis set. The input or expected values are plotted as vertical dashed black lines in the histogram plots. Each basis set manages to fit the spectrum's light and recover a reasonable star formation history. But, the DFK-AVG and DFK-SSP results are more precise as evidenced by their narrower histograms.}
\label{fig:tau2gyr-dtm}
\end{figure*}

The results from using \textsc{sspmodel} with the DFK-AVG, TB, and DFK-SSP basis sets for the low signal-to-noise, early-type ($\tau=2$~Gyr) synthetic galaxies are summarized in Fig. \ref{fig:tau2gyr-dtm} and Table \ref{tab:rctable}. In particular note the ``Mean Frac'l Error" and ``95 Percentile" columns in Table \ref{tab:rctable}. Definitions for these two columns are found in the caption of Table \ref{tab:rctable}. Together these quickly indicate the accuracy and precision, both expressed in terms of fractional deviations from the true input model value of the parameter in question. Our major findings for the early-type galaxy tests are: (1) the TB, DFK-AVG, and  DFK-SSP basis sets can all reproduce the input galaxy spectrum well as all the basis sets have very similar medians and widths of their distributions of  $\chi_{\nu}^{2}$; (2) all basis sets tested recovered a mean $A_{V}$ to within 5 per cent of its input value; (3) though the DFK-AVG and DFK-SSP basis sets recovered systematically different mass-to-light ratios than the input and the TB basis set, this is easily understood, and their results are more precise; lastly, (4) each basis set recovered reasonable mean mass fractions for the Y and I1 age bins given their small light and mass fractions and accurate mean mass fractions for the O age bin.

To understand the ability of the various basis sets to recover $M/L_{V}$, it is important to consider how the basis sets were formed. The DFK-AVG and DFK-SSP basis set results are offset from the expected value of $M/L_{V}$ by about 16 per cent and 13 per cent, respectively, but the DFK-AVG and DFK-SSP basis sets were more precise (narrower histograms) than the TB basis set in their reported values of $M/L_{V}$. We can understand this offset by looking at Fig. \ref{fig:sfhs}. For the largely old and steeply peaked star formation history of the synthetic early-type galaxy ($\tau=$2~Gyr), most of the star formation and thus light in the spectrum comes from stellar populations in the O age bin. To fit the light from this old stellar population with the DFK-AVG basis set, \textsc{sspmodel} will primarily use the O base. The O base, however, is an average that includes stellar populations younger than the vast majority of stars comprising the input spectrum that will contribute more light for their mass. Therefore, to match the input spectrum's light, \textsc{sspmodel} will find it needs less of the O base (less mass in the O age bin). Consequently, the mass-to-light ratio recovered by the DFK-AVG basis set will be systematically low. Regarding precision, recovered $M/L_{V}$ values from the TB basis set vary by as much as 13 per cent of the mean recovered $M/L_{V}$ value 95 per cent of the time; whereas for the DFK-AVG basis, though the results are offset, the recovered $M/L_{V}$ values only vary by at most $\sim$2 per cent of the mean recovered $M/L_{V}$ value. In fact, for the 6 parameters recovered ($A_{V}$, $M/L_{V}$, Y, I1, I2, O), the DFK-AVG basis set has the most precise results in all but the I1 age bin for this star formation history. This precision of recovered parameters can be an attractive property when analyzing single galaxy spectra provided that any systematics can be corrected. In practice, though, the systematic offset we see in the mass-to-light ratio is small compared to the uncertainty in the mass-to-light ratios due to metallicity, abundance ratios, and the initial mass function \citep[cf.][]{Smith-2014}.

\begin{figure*}
\includegraphics{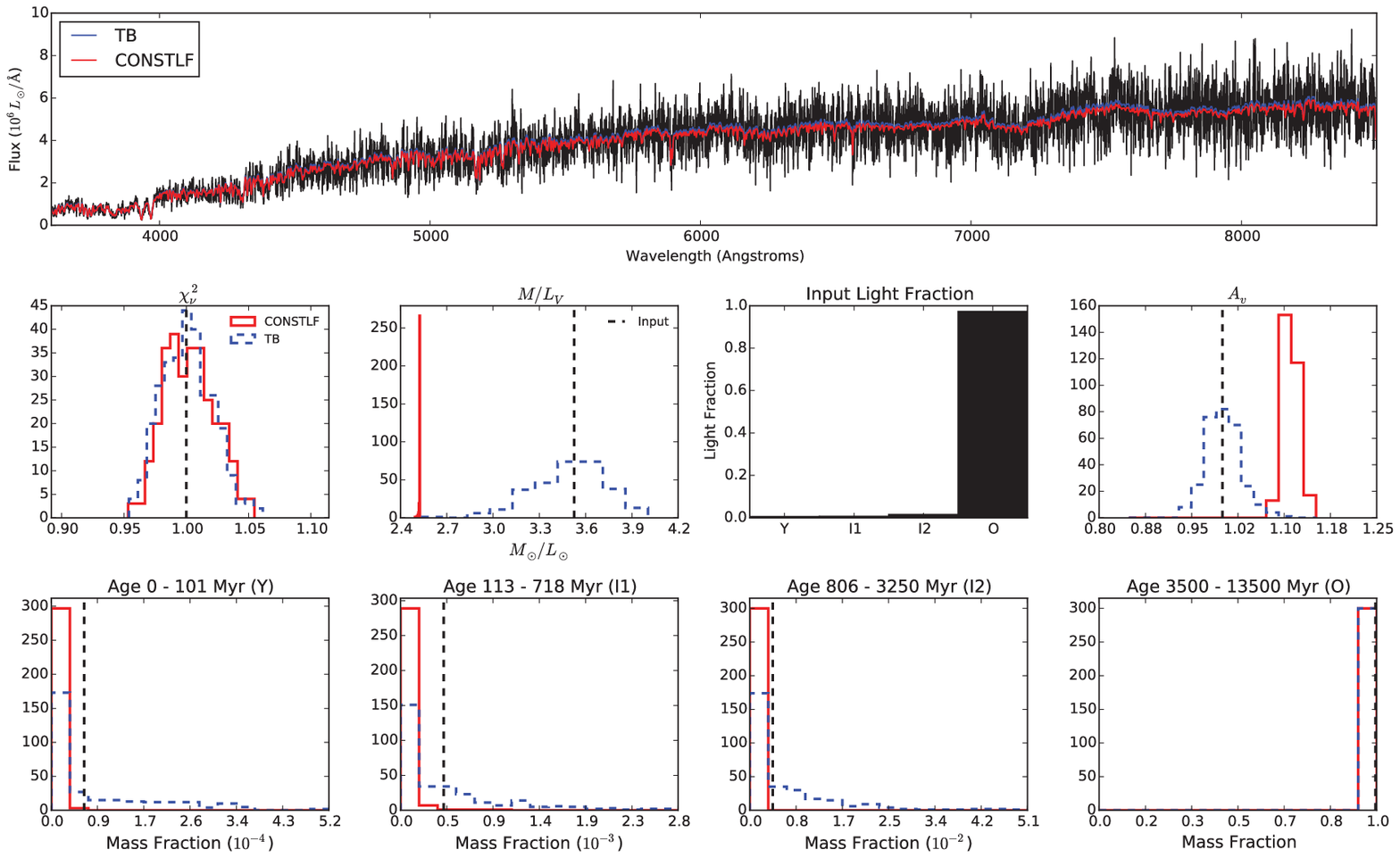}
\caption{The results of running \textsc{sspmodel} on 300 realizations of a synthetic late-type ($\tau=$ 10~Gyr) galaxy spectrum with S/N$\sim$5~\AA$^{-1}$ for the DFK-AVG, TB, and DFK-SSP basis sets. See Fig. \ref{fig:tau2gyr-dtm} for plot description. Note that all basis sets reproduce the input noisy spectrum, but the DFK-AVG basis set consistently recovers accurate and precise parameters. }
\label{fig:tau10gyr-dtm}
\end{figure*}

A careful reader might note in conclusion (4), we omit the I2 age bin. This is because each basis set has trouble recovering an accurate mass fraction even though the light fraction estimated for this population is about 7 per cent. Note that in Fig. \ref{fig:tau2gyr-dtm}, the DFK-AVG and the DFK-SSP I2 histograms appear to find the correct mean mass fraction, but this is merely due to the coarse binning required to show the more discrepant values recovered by the TB basis set. Each basis set's mean mass fraction for the stellar populations in age bin I2 is $>$50 per cent away from the expected value (see Table \ref{tab:rctable}). These results suggest that populations in the I2 age bin contributing $<$ 7 per cent of the light in low signal-to-noise spectra are not likely to be recovered well. Note also in Fig. \ref{fig:tau2gyr-dtm} that the presence of discrepant values that form the tails of the TB basis set's distributions of recovered mass fractions in age bins I2 and O are indicative that the TB basis set is having more trouble selecting which stellar populations to use (those of the I2 or O age bin). As a result, the TB basis set trades off between the two age bins. 

To summarize, we find that for a typical early-type galaxy spectrum each basis set tested can reproduce the noisy input spectrum and adequately recover host galaxy $A_{V}$. The TB basis set more accurately determines the mass-to-light ratio of the spectrum--though with more scatter--than the DFK-AVG and DFK-SSP basis sets, but we understand this systematic offset. We also find each basis set recovers reasonable mean mass fractions of the youngest stellar populations in age bins Y and I1 and accurate mean mass fractions of the oldest stellar populations in age bin O that account for the majority of the light. And in general, we find that the DFK-AVG basis set recovers more precise parameter values.

\subsection{Tau = 10 Gyr Model Results}\label{tau10}
The results of the DFK-AVG, TB, and DFK-SSP basis sets on the low signal-to-noise late-type synthetic galaxy spectra ($\tau = 10$~Gyr) are summarized in Fig. \ref{fig:tau10gyr-dtm} and Table \ref{tab:rctable}. The main findings are that (1) each basis set was able to reproduce the input spectrum and recover an accurate (within 2 per cent of the input) mean for the host galaxy $A_{V}$, (2) all basis sets except the DFK-SSP basis set recovered an accurate mean mass-to-light ratio, and (3) the DFK-AVG and TB basis sets recovered more accurate star formation histories than the DFK-SSP basis set with the DFK-AVG basis set recovering the more precise and accurate star formation history.

With regard to the mass-to-light ratio, the mean values of the recovered mass-to-light ratio of the DFK-AVG and TB basis set runs were accurate to within 10 per cent of the expected value for $M/L_{V}$, but the DFK-SSP basis set's mean value of $M/L_{V}$ was only accurate to within 25 per cent.  Since the $\tau = 10$~Gyr star formation rate changes slowly with time across all of the age bins defined by the DFK-AVG basis set (see Fig. \ref{fig:sfhs}), our implicit assumption of constant star formation in the formation of our DFK-AVG basis set is reasonable. As a result, unlike in the early-type galaxy tests, the mass-to-light ratio is recovered well by the DFK-AVG basis set. Similar to the results in the early-type galaxy star formation history tests, the DFK-AVG basis set recovered more precise $M/L_{V}$ values than the TB basis set. 

In recovering the late-type galaxy star formation history, each basis set recovered a consistent mean mass fraction of the youngest stellar populations in age bin Y, though the DFK-AVG basis set recovered the most accurate mean value. Fig. \ref{fig:tau10gyr-dtm} shows that the mean mass fractions recovered for the I1,I2, and O age bins are fairly accurate for the DFK-AVG and TB basis sets with the more accurate and precise means belonging to the DFK-AVG basis set. Table \ref{tab:rctable} more precisely shows that the DFK-AVG basis set mean mass fractions for these age bins are within 10 per cent of the expected values. The mean mass fractions recovered by the TB basis set are within 49 per cent of the expected values. As explained above, the $\tau=10$~Gyr star formation history is well matched to the DFK-AVG basis set, so it is not surprising that the DFK-AVG basis set performs well. But, the DFK-SSP basis set recovered the least accurate mean mass fractions for the I1, I2, and O age bins. Since the DFK-SSP basis set is a set of SSPs, or 4 discrete bursts, it is not well matched to the approximately constant star formation history of this $\tau = 10$~Gyr test. This is likely why the mean mass fractions recovered by the DFK-SSP basis set are not as accurate.

In summary, we find that for a typical late-type galaxy spectrum, each basis set was able to reproduce the noisy input galaxy spectrum and recover host galaxy $A_V$. The DFK-AVG and TB basis sets do a better job at recovering an accurate mass-to-light ratio and the star formation history in the broad age bins. Specifically, the DFK-AVG basis set is generally more accurate (in each case of recovered parameters) and precise (in all but the $A_{V}$ and Y mass fraction). The DFK-SSP basis set, however, had more trouble than the DFK-AVG and TB basis sets in recovering the mass-to-light ratio and the star formation history. We require a basis set that has at least the versatility of fitting both early and late-type galaxy spectra, so in light of these results, we exclude the DFK-SSP basis set from further analysis.

\begin{figure*}
\includegraphics{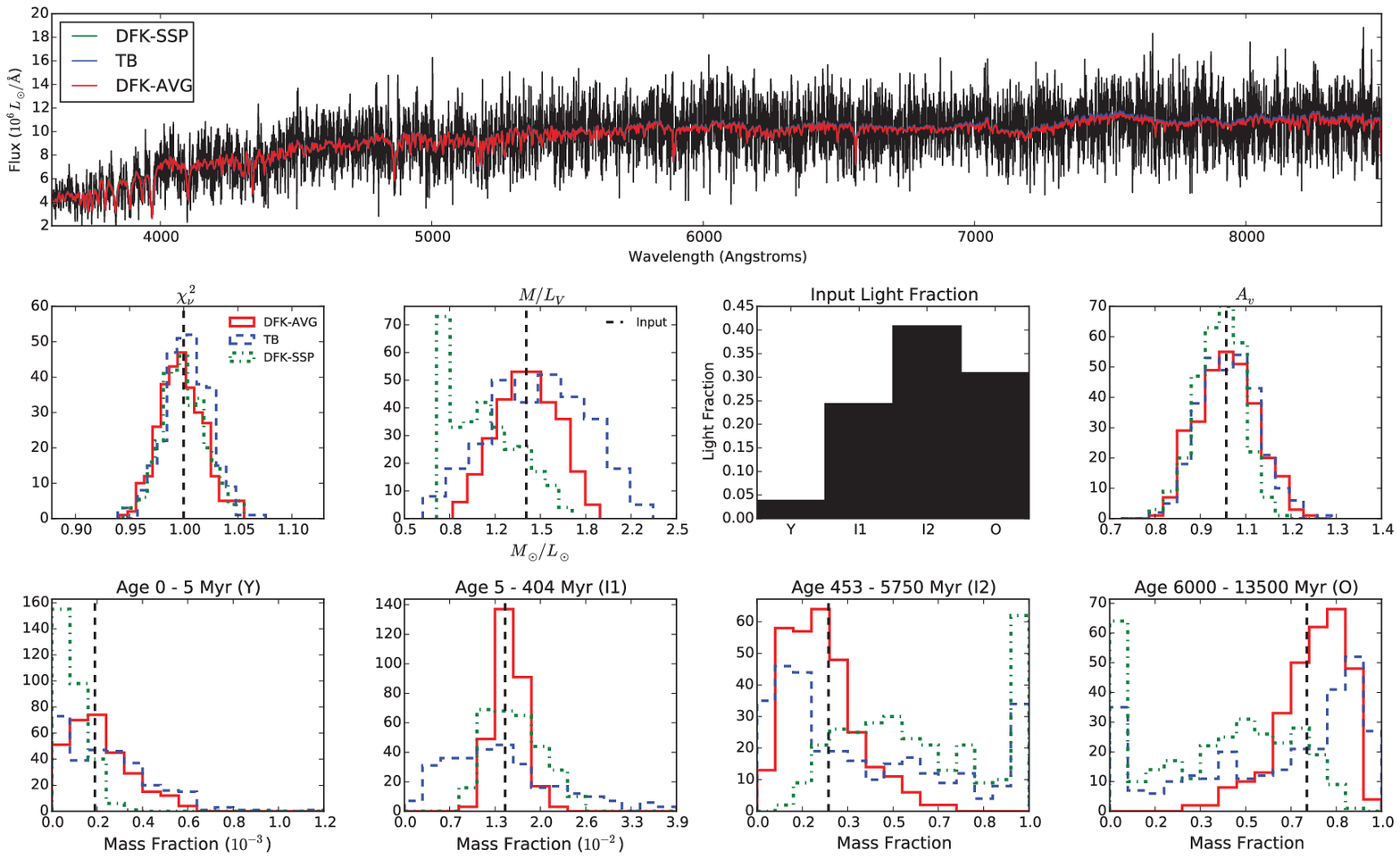}
\caption{The results of running \textsc{sspmodel} on 300 realizations of an synthetic early-type ($\tau=$ 2~Gyr) galaxy spectrum with S/N$\sim$5~\AA$^{-1}$ for the CONSTLF and TB basis sets. See Fig. \ref{fig:tau2gyr-dtm} for plot description. Note that all basis sets reproduce the input noisy spectrum.}
\label{fig:tau2gyr-ct}
\end{figure*}

\begin{figure*}
\includegraphics{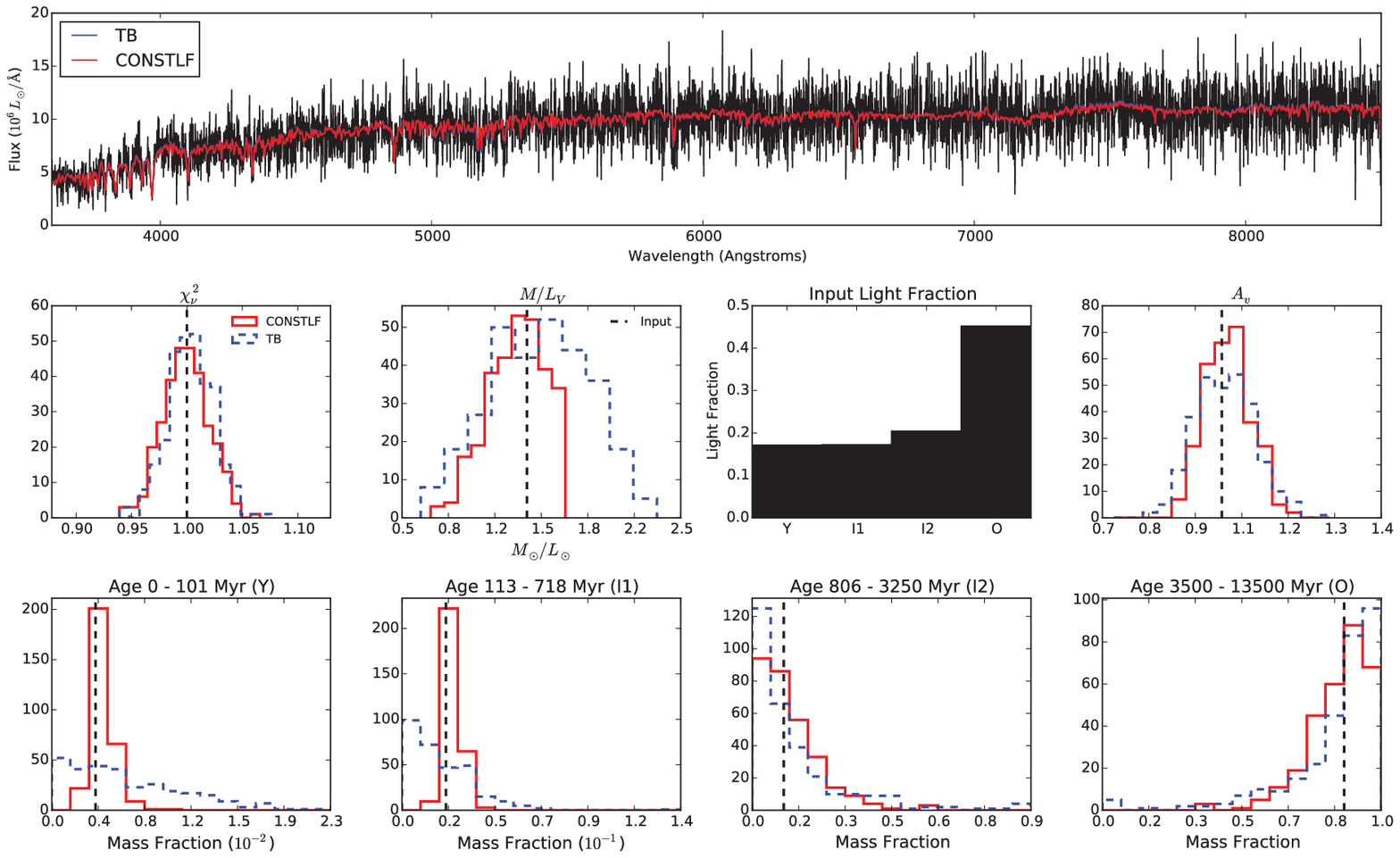}
\caption{The results of running \textsc{sspmodel} on 300 realizations of a synthetic late-type ($\tau=$ 10~Gyr) galaxy spectrum with S/N$\sim$5~\AA$^{-1}$ for the CONSTLF and TB basis sets. See Fig. \ref{fig:tau2gyr-dtm} for plot description. Note that both basis sets reproduce the input noisy spectrum.}
\label{fig:tau10gyr-ct}
\end{figure*}

\subsection{Constant Light Fraction Basis Set Results} \label{constlf-res}
The results of the CONSTLF basis set on the low signal-to-noise early-type and late-type galaxies are tabulated in Table \ref{tab:constlf-rctable} and represented graphically in Fig. \ref{fig:tau2gyr-ct} and \ref{fig:tau10gyr-ct}. These results are summarized graphically in Fig. \ref{fig:rcfig} alongside the results for each star formation history tested. The constant light fraction basis set, CONSTLF, cannot be binned the same as the DFK-AVG basis sets, so we analysed its results separately. Similar to the other 3 basis sets tested, the CONSTLF basis set can reproduce the spectrum of a  synthetic galaxy with a $\tau$ = 2 or 10~Gyr star formation history and does so with similar precision as determined from its distribution of $\chi_{\nu}^2$. But, when we compare the accuracy of the CONSTLF basis set to the accuracy of the DFK-AVG and TB basis sets using the mean fractional error in Table \ref{tab:rctable} and Table \ref{tab:constlf-rctable} it is generally less accurate. Given these results, we also exclude the CONSTLF basis set from further tests.

\begin{sidewaystable*}\centering
\caption{Table of results from using the DFK-AVG, TB, and DFK-SSP basis sets to analyze synthetic galaxies for 6 star formation histories. The data is arranged by star formation history vertically, and by basis set horizontally. Under each basis set heading, there are 4 columns: (1) mean value recovered by the basis set out of the 300 runs, (2) median value recovered, (3) the fractional error of the mean value recovered (i.e. the difference of the mean and the true value divided by the true value), and (4) the range of \textit{fractional error values}, not parameter values, containing the central 95\% (about the mean) of the 300 recovered values for each parameter. Column (3) measures the accuracy of a basis set. Column (4) measures the precision. The DFK-AVG basis set is quite often the most precise. For the 6 parameters of interest ($A_{V}$, $M/L_{V}$, Y, I1, I2, O) for the 6 star formation histories below, DFK-AVG is the most precise basis set 33 out of 36 times. Regarding accuracy, the DFK-AVG and TB basis sets are closer to being evenly matched. DFK-AVG has the most accurate mean 17 out of 36 times. The TB basis set has the most accurate mean 16/36 times.}
\resizebox{\linewidth}{!}{
\begin{tabular}{@{}ccccccccccccccc@{}}\toprule
&  \multicolumn{2}{c}{Input}  &  \multicolumn{4}{c}{DFK-AVG} &  \multicolumn{4}{c}{TB} &  \multicolumn{4}{c}{DFK-SSP}\\
\cmidrule(l{30pt}r{30pt}){2-3}
\cmidrule(l{5pt}r{5pt}){4-7}
\cmidrule(l{5pt}r{5pt}){8-11}
\cmidrule(l{5pt}r{5pt}){12-15}
\multicolumn{3}{c}{} & Mean & Median & Mean Frac'l Error & 95 Percentile & Mean & Median & Mean Frac'l Error & 95 Percentile & Mean & Median & Mean Frac'l Error & 95 Percentile \\
\cmidrule(l{5pt}r{5pt}){4-4}
\cmidrule(l{5pt}r{5pt}){5-5}
\cmidrule(l{5pt}r{5pt}){6-6}
\cmidrule(l{5pt}r{5pt}){7-7}
\cmidrule(l{5pt}r{5pt}){8-8}
\cmidrule(l{5pt}r{5pt}){9-9}
\cmidrule(l{5pt}r{5pt}){10-10}
\cmidrule(l{5pt}r{5pt}){11-11}
\cmidrule(l{5pt}r{5pt}){12-12}
\cmidrule(l{5pt}r{5pt}){13-13}
\cmidrule(l{5pt}r{5pt}){14-14}
\cmidrule(l{5pt}r{5pt}){15-15}
& $A_V$ & 1.000 & 1.050 & 1.051 & \textbf{0.050} & \textbf{(0.012, 0.083)}  & 1.002 & 0.999 & \textbf{0.002} & \textbf{(-0.049, 0.066)}  & 1.019 & 1.019 & \textbf{0.019} & \textbf{(-0.017, 0.064)} \\
\multicolumn{1}{c}{Tau 2 Gyr} & $M/L_{V}$ & 3.526 & 2.972 & 2.981 & \textbf{-0.157} & \textbf{(-0.177, -0.148)}  & 3.499 & 3.531 & \textbf{-0.008} & \textbf{(-0.140, 0.102)}  & 3.067 & 3.087 & \textbf{-0.130} & \textbf{(-0.173, -0.117)} \\
& $\chi_{\nu}^{2}$ & 1.000 & 1.000 & 1.000 & \textbf{-0.000} & \textbf{(-0.035, 0.040)}  & 1.001 & 1.000 & \textbf{0.001} & \textbf{(-0.035, 0.043)}  & 1.000 & 1.000 & \textbf{-0.000} & \textbf{(-0.040, 0.039)} \\
\\
& Light Fraction & Mass Fraction & \\
\cmidrule(l{5pt}r{5pt}){2-2}
\cmidrule(l{5pt}r{5pt}){3-3}
\multicolumn{1}{r}{Y} & 0.001 & 0.0 & 0.000 & 0.000 & \textbf{-0.211} & \textbf{(-1.000, 3.011)}  & 0.000 & 0.000 & \textbf{0.804} & \textbf{(-0.998, 4.328)}  & 0.000 & 0.000 & \textbf{-0.162} & \textbf{(-1.000, 3.140)} \\
\multicolumn{1}{r}{I1} & 0.01 & 0.0 & 0.000 & 0.000 & \textbf{-0.382} & \textbf{(-1.000, 0.539)}  & 0.000 & 0.000 & \textbf{-0.242} & \textbf{(-0.999, 2.013)}  & 0.000 & 0.000 & \textbf{-0.369} & \textbf{(-1.000, 0.199)} \\
\multicolumn{1}{r}{I2} & 0.067 & 0.019 & 0.002 & 0.000 & \textbf{-0.895} & \textbf{(-1.000, -0.292)}  & 0.055 & 0.031 & \textbf{1.848} & \textbf{(-0.998, 10.349)}  & 0.005 & 0.000 & \textbf{-0.744} & \textbf{(-1.000, 0.698)} \\
\multicolumn{1}{r}{O} & 0.922 & 0.98 & 0.998 & 1.000 & \textbf{0.018} & \textbf{(0.006, 0.020)}  & 0.945 & 0.969 & \textbf{-0.036} & \textbf{(-0.204, 0.019)}  & 0.995 & 1.000 & \textbf{0.015} & \textbf{(-0.014, 0.020)} \\
\midrule
& $A_V$ & 1.000 & 1.004 & 1.002 & \textbf{0.004} & \textbf{(-0.127, 0.142)}  & 1.008 & 1.008 & \textbf{0.008} & \textbf{(-0.114, 0.153)}  & 0.982 & 0.987 & \textbf{-0.018} & \textbf{(-0.137, 0.091)} \\
\multicolumn{1}{c}{Tau 10 Gyr} & $M/L_{V}$ & 1.397 & 1.401 & 1.399 & \textbf{0.003} & \textbf{(-0.304, 0.281)}  & 1.488 & 1.507 & \textbf{0.065} & \textbf{(-0.425, 0.540)}  & 1.083 & 1.072 & \textbf{-0.225} & \textbf{(-0.458, 0.144)} \\
& $\chi_{\nu}^{2}$ & 1.000 & 0.998 & 0.997 & \textbf{-0.002} & \textbf{(-0.041, 0.043)}  & 1.004 & 1.004 & \textbf{0.004} & \textbf{(-0.037, 0.042)}  & 1.000 & 0.999 & \textbf{-0.000} & \textbf{(-0.040, 0.042)} \\
\\
& Light Fraction & Mass Fraction & \\
\cmidrule(l{5pt}r{5pt}){2-2}
\cmidrule(l{5pt}r{5pt}){3-3}
\multicolumn{1}{r}{Y} & 0.039 & 0.0 & 0.000 & 0.000 & \textbf{0.110} & \textbf{(-1.000, 1.758)}  & 0.000 & 0.000 & \textbf{0.292} & \textbf{(-0.999, 2.296)}  & 0.000 & 0.000 & \textbf{-0.549} & \textbf{(-1.000, 0.249)} \\
\multicolumn{1}{r}{I1} & 0.244 & 0.014 & 0.015 & 0.015 & \textbf{0.037} & \textbf{(-0.225, 0.342)}  & 0.014 & 0.013 & \textbf{-0.042} & \textbf{(-0.819, 1.354)}  & 0.016 & 0.015 & \textbf{0.098} & \textbf{(-0.314, 0.666)} \\
\multicolumn{1}{r}{I2} & 0.408 & 0.26 & 0.237 & 0.220 & \textbf{-0.089} & \textbf{(-0.774, 1.075)}  & 0.387 & 0.278 & \textbf{0.487} & \textbf{(-0.854, 2.765)}  & 0.594 & 0.545 & \textbf{1.284} & \textbf{(-0.319, 2.772)} \\
\multicolumn{1}{r}{O} & 0.309 & 0.725 & 0.748 & 0.763 & \textbf{0.031} & \textbf{(-0.389, 0.277)}  & 0.599 & 0.712 & \textbf{-0.174} & \textbf{(-0.998, 0.311)}  & 0.390 & 0.441 & \textbf{-0.463} & \textbf{(-1.000, 0.121)} \\
\midrule
& $A_V$ & 1.000 & 0.919 & 0.919 & \textbf{-0.081} & \textbf{(-0.196, 0.024)}  & 0.961 & 0.963 & \textbf{-0.039} & \textbf{(-0.306, 0.243)} \\
\multicolumn{1}{c}{Tau 2 Gyr 100 Myr Burst (Peak)} & $M/L_{V}$ & 0.412 & 0.970 & 0.970 & \textbf{1.355} & \textbf{(1.076, 1.629)}  & 0.365 & 0.318 & \textbf{-0.115} & \textbf{(-0.856, 1.104)} \\
& $\chi_{\nu}^{2}$ & 1.000 & 1.006 & 1.006 & \textbf{0.006} & \textbf{(-0.033, 0.046)}  & 1.002 & 1.003 & \textbf{0.002} & \textbf{(-0.039, 0.045)} \\
\\
& Light Fraction & Mass Fraction & \\
\cmidrule(l{5pt}r{5pt}){2-2}
\cmidrule(l{5pt}r{5pt}){3-3}
\multicolumn{1}{r}{Y} & 0.458 & 0.005 & 0.004 & 0.004 & \textbf{-0.115} & \textbf{(-0.261, 0.032)}  & 0.012 & 0.007 & \textbf{1.628} & \textbf{(-0.552, 10.122)} \\
\multicolumn{1}{r}{I1} & 0.341 & 0.043 & 0.014 & 0.014 & \textbf{-0.680} & \textbf{(-0.840, -0.490)}  & 0.101 & 0.056 & \textbf{1.352} & \textbf{(-0.733, 10.261)} \\
\multicolumn{1}{r}{I2} & 0.014 & 0.018 & 0.000 & 0.000 & \textbf{-1.000} & \textbf{(-1.000, -1.000)}  & 0.309 & 0.109 & \textbf{15.735} & \textbf{(-0.987, 49.025)} \\
\multicolumn{1}{r}{O} & 0.188 & 0.934 & 0.982 & 0.982 & \textbf{0.052} & \textbf{(0.043, 0.059)}  & 0.578 & 0.773 & \textbf{-0.381} & \textbf{(-0.999, 0.048)} \\
\midrule
& $A_V$ & 1.000 & 0.988 & 0.987 & \textbf{-0.012} & \textbf{(-0.136, 0.114)}  & 0.997 & 0.995 & \textbf{-0.003} & \textbf{(-0.276, 0.282)} \\
\multicolumn{1}{c}{Tau 10 Gyr 100 Myr Burst (Peak)} & $M/L_{V}$ & 0.363 & 0.899 & 0.894 & \textbf{1.477} & \textbf{(1.110, 1.838)}  & 0.454 & 0.416 & \textbf{0.249} & \textbf{(-0.718, 1.948)} \\
& $\chi_{\nu}^{2}$ & 1.000 & 1.004 & 1.004 & \textbf{0.004} & \textbf{(-0.032, 0.041)}  & 1.002 & 1.000 & \textbf{0.002} & \textbf{(-0.038, 0.042)} \\
\\
& Light Fraction & Mass Fraction & \\
\cmidrule(l{5pt}r{5pt}){2-2}
\cmidrule(l{5pt}r{5pt}){3-3}
\multicolumn{1}{r}{Y} & 0.383 & 0.005 & 0.004 & 0.004 & \textbf{-0.198} & \textbf{(-0.352, -0.013)}  & 0.008 & 0.005 & \textbf{0.533} & \textbf{(-0.572, 4.372)} \\
\multicolumn{1}{r}{I1} & 0.361 & 0.057 & 0.023 & 0.023 & \textbf{-0.591} & \textbf{(-0.736, -0.422)}  & 0.073 & 0.050 & \textbf{0.269} & \textbf{(-0.752, 3.308)} \\
\multicolumn{1}{r}{I2} & 0.146 & 0.248 & 0.000 & 0.000 & \textbf{-1.000} & \textbf{(-1.000, -1.000)}  & 0.335 & 0.174 & \textbf{0.353} & \textbf{(-0.999, 2.769)} \\
\multicolumn{1}{r}{O} & 0.11 & 0.69 & 0.973 & 0.973 & \textbf{0.409} & \textbf{(0.395, 0.422)}  & 0.585 & 0.748 & \textbf{-0.153} & \textbf{(-0.999, 0.416)} \\
\midrule
& $A_V$ & 1.000 & 0.996 & 1.000 & \textbf{-0.004} & \textbf{(-0.163, 0.163)}  & 0.992 & 0.983 & \textbf{-0.008} & \textbf{(-0.185, 0.212)} \\
\multicolumn{1}{c}{Tau 2 Gyr 100 Myr Burst (After Peak)} & $M/L_{V}$ & 0.624 & 0.430 & 0.421 & \textbf{-0.310} & \textbf{(-0.663, 0.041)}  & 0.576 & 0.589 & \textbf{-0.076} & \textbf{(-0.760, 0.518)} \\
& $\chi_{\nu}^{2}$ & 1.000 & 0.999 & 1.000 & \textbf{-0.001} & \textbf{(-0.040, 0.037)}  & 1.002 & 1.002 & \textbf{0.002} & \textbf{(-0.038, 0.039)} \\
\\
& Light Fraction & Mass Fraction & \\
\cmidrule(l{5pt}r{5pt}){2-2}
\cmidrule(l{5pt}r{5pt}){3-3}
\multicolumn{1}{r}{Y} & 0.034 & 0.0 & 0.000 & 0.000 & \textbf{-0.470} & \textbf{(-1.000, 2.351)}  & 0.001 & 0.000 & \textbf{0.705} & \textbf{(-1.000, 4.222)} \\
\multicolumn{1}{r}{I1} & 0.761 & 0.09 & 0.202 & 0.185 & \textbf{1.239} & \textbf{(0.166, 3.673)}  & 0.135 & 0.095 & \textbf{0.491} & \textbf{(-0.406, 4.621)} \\
\multicolumn{1}{r}{I2} & 0.014 & 0.018 & 0.024 & 0.000 & \textbf{0.354} & \textbf{(-1.000, 13.207)}  & 0.101 & 0.020 & \textbf{4.725} & \textbf{(-0.995, 37.304)} \\
\multicolumn{1}{r}{O} & 0.191 & 0.892 & 0.774 & 0.809 & \textbf{-0.132} & \textbf{(-0.572, 0.001)}  & 0.764 & 0.869 & \textbf{-0.143} & \textbf{(-0.966, 0.059)} \\
\midrule
& $A_V$ & 1.000 & 1.011 & 1.012 & \textbf{0.011} & \textbf{(-0.177, 0.186)}  & 1.009 & 1.001 & \textbf{0.009} & \textbf{(-0.195, 0.239)} \\
\multicolumn{1}{c}{Tau 10 Gyr 100 Myr Burst (After Peak)} & $M/L_{V}$ & 0.516 & 0.445 & 0.443 & \textbf{-0.139} & \textbf{(-0.621, 0.431)}  & 0.578 & 0.579 & \textbf{0.120} & \textbf{(-0.679, 0.987)} \\
& $\chi_{\nu}^{2}$ & 1.000 & 1.001 & 1.000 & \textbf{0.001} & \textbf{(-0.042, 0.043)}  & 1.001 & 1.002 & \textbf{0.001} & \textbf{(-0.039, 0.044)} \\
\\
& Light Fraction & Mass Fraction & \\
\cmidrule(l{5pt}r{5pt}){2-2}
\cmidrule(l{5pt}r{5pt}){3-3}
\multicolumn{1}{r}{Y} & 0.041 & 0.0 & 0.000 & 0.000 & \textbf{-0.586} & \textbf{(-1.000, 1.252)}  & 0.001 & 0.001 & \textbf{0.765} & \textbf{(-0.999, 5.899)} \\
\multicolumn{1}{r}{I1} & 0.699 & 0.104 & 0.189 & 0.165 & \textbf{0.812} & \textbf{(-0.148, 3.375)}  & 0.135 & 0.090 & \textbf{0.290} & \textbf{(-0.551, 3.739)} \\
\multicolumn{1}{r}{I2} & 0.148 & 0.236 & 0.179 & 0.105 & \textbf{-0.241} & \textbf{(-1.000, 2.064)}  & 0.250 & 0.119 & \textbf{0.059} & \textbf{(-0.999, 2.602)} \\
\multicolumn{1}{r}{O} & 0.112 & 0.659 & 0.631 & 0.725 & \textbf{-0.042} & \textbf{(-1.000, 0.365)}  & 0.614 & 0.792 & \textbf{-0.068} & \textbf{(-0.998, 0.424)} \\
\bottomrule
\end{tabular}
}

\label{tab:rctable}
\end{sidewaystable*}
\begin{table*}\centering
\caption{Table of results from using the CONSTLF basis set for the 2 $\tau$-model star formation histories in \S{\ref{testedsfhs}}. This table has the same format as Table \ref{tab:rctable}. Columns ``Mean Fract'l Error" and ``95 Percentile" are represented graphically in Figure \ref{fig:rcfig}. In general, the CONSTLF basis set can reproduce the input spectrum, but its accuracy and precision are not often better than the other 3 basis sets tested.}
\resizebox{\linewidth}{!}{
\begin{tabular}{@{}ccccccc@{}}\toprule
&  \multicolumn{2}{c}{Input}  &  \multicolumn{4}{c}{CONSTLF}\\
\cmidrule(l{30pt}r{30pt}){2-3}
\cmidrule(l{5pt}r{5pt}){4-7}
\multicolumn{3}{c}{} & Mean & Median & Mean Frac'l Error & 95 Percentile \\
\cmidrule(l{5pt}r{5pt}){4-4}
\cmidrule(l{5pt}r{5pt}){5-5}
\cmidrule(l{5pt}r{5pt}){6-6}
\cmidrule(l{5pt}r{5pt}){7-7}
& $A_V$ & 1.000 & 1.111 & 1.109 & \textbf{0.111} & \textbf{(0.088, 0.135)} \\
\multicolumn{1}{c}{Tau 2 Gyr} & $M/L_{V}$ & 3.526 & 2.525 & 2.526 & \textbf{-0.284} & \textbf{(-0.286, -0.284)} \\
& $\chi_{\nu}^{2}$ & 1.000 & 1.003 & 1.002 & \textbf{0.003} & \textbf{(-0.032, 0.042)} \\
\\
& Light Fraction & Mass Fraction & \\
\cmidrule(l{5pt}r{5pt}){2-2}
\cmidrule(l{5pt}r{5pt}){3-3}
\multicolumn{1}{r}{Y} & 0.005 & 0.0 & 0.000 & 0.000 & \textbf{-0.977} & \textbf{(-1.000, -0.674)} \\
\multicolumn{1}{r}{I1} & 0.006 & 0.0 & 0.000 & 0.000 & \textbf{-0.947} & \textbf{(-1.000, -0.393)} \\
\multicolumn{1}{r}{I2} & 0.015 & 0.004 & 0.000 & 0.000 & \textbf{-1.000} & \textbf{(-1.000, -1.000)} \\
\multicolumn{1}{r}{O} & 0.973 & 0.995 & 1.000 & 1.000 & \textbf{0.005} & \textbf{(0.004, 0.005)} \\
\midrule
& $A_V$ & 1.000 & 1.014 & 1.014 & \textbf{0.014} & \textbf{(-0.088, 0.123)} \\
\multicolumn{1}{c}{Tau 10 Gyr} & $M/L_{V}$ & 1.397 & 1.319 & 1.342 & \textbf{-0.056} & \textbf{(-0.339, 0.168)} \\
& $\chi_{\nu}^{2}$ & 1.000 & 0.999 & 0.999 & \textbf{-0.001} & \textbf{(-0.041, 0.039)} \\
\\
& Light Fraction & Mass Fraction & \\
\cmidrule(l{5pt}r{5pt}){2-2}
\cmidrule(l{5pt}r{5pt}){3-3}
\multicolumn{1}{r}{Y} & 0.171 & 0.004 & 0.004 & 0.004 & \textbf{0.153} & \textbf{(-0.185, 0.717)} \\
\multicolumn{1}{r}{I1} & 0.172 & 0.022 & 0.026 & 0.026 & \textbf{0.154} & \textbf{(-0.163, 0.578)} \\
\multicolumn{1}{r}{I2} & 0.204 & 0.107 & 0.122 & 0.094 & \textbf{0.144} & \textbf{(-1.000, 2.474)} \\
\multicolumn{1}{r}{O} & 0.452 & 0.867 & 0.848 & 0.877 & \textbf{-0.022} & \textbf{(-0.315, 0.122)} \\
\bottomrule
\end{tabular}
}

\label{tab:constlf-rctable}
\end{table*}

\subsection{Burst Model Testing Results}\label{tauburst}
\begin{figure*}
\includegraphics{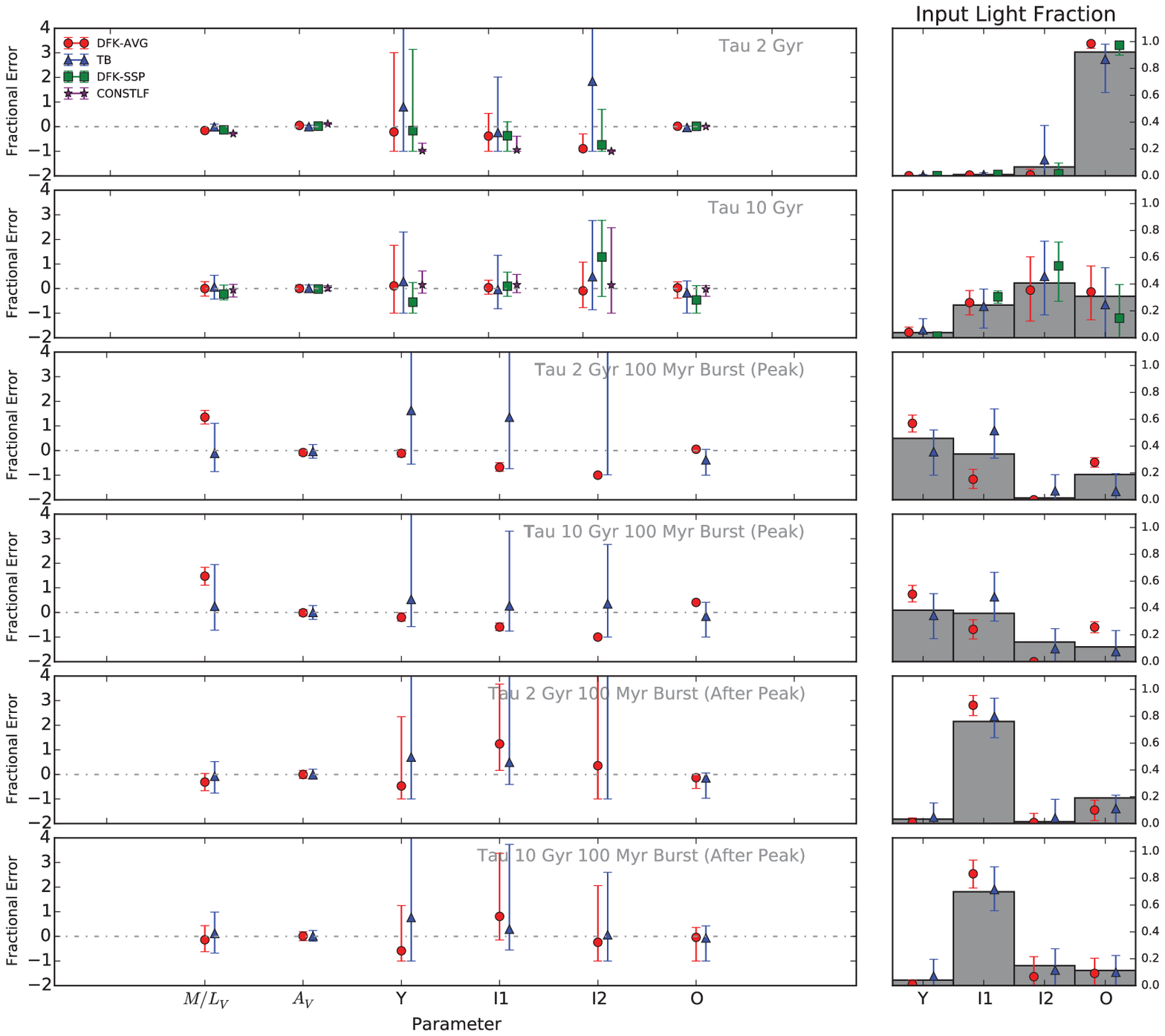}
\caption{The results of \textsc{sspmodel} for the DFK, TB, DFK-SSP, and CONSTLF basis sets for the 6 star formation histories tested. The fractional error of the mean (i.e. the difference of the mean and the true value divided by the true value) of $M/L_{V}$, $A_{V}$, and the mass fractions in the 4 broad age bins defined in Table \ref{tab:test-bases-ages} are shown with 95 percentile intervals as error bars. The error bars show the range of fractional errors that 95 per cent of the recovered values have for each basis set. A value of 1 on the vertical scale of these plots represents a 100 per cent deviation from the true input quantity. Corresponding numerical values are given in Tables  \ref{tab:rctable} and \ref{tab:constlf-rctable} (columns ``Mean Fract'l Error" and ``95 Percentile" ). The input star formation history's light fraction in these broad age bins is shown on the right panels with the mean values of each basis set plotted as points with error bars that show the range of values that 95 per cent of the recovered values have for each basis set (CONSTLF is not included as it has different age bins). Note there are two TB points missing in the third and fifth rows on the left because the mean fractional errors lie outside the plotting limits. The DFK-AVG basis set's results are often the most precise (small error bars) and just edge out the TB basis set in having more accurate means (closer to zero axis). On the right, the DFK-AVG basis set does a noticeably better job of resolving the burst in galaxies observed during a peak in star formation.}
\label{fig:rcfig}
\end{figure*}

The results of the TB and DFK-AVG basis sets on the low signal-to-noise synthetic galaxies with the 4 burst star formation histories described in the beginning of \S\ref{testedsfhs} are summarized in Table \ref{tab:rctable} and in Fig. \ref{fig:rcfig}. For the basis testing with the bursts, we do not test the DFK-SSP and CONSTLF basis sets for the reasons previously stated. 

For all of the burst star formation histories, we find that both the DFK-AVG and TB basis sets can recover means of the host galaxy $A_{V}$ within 10 per cent of its input value. The DFK-AVG basis set has the most precise parameter values recovered for the primary 6 parameters for all burst star formation histories, but Fig. \ref{fig:rcfig} and Table \ref{tab:rctable} show that the DFK-AVG and TB basis sets have more accurate mean mass fractions in different star formation histories in different age bins. For the early type galaxy observed at the burst peak, the DFK-AVG basis set has more accurate mean mass fractions in all age bins. The DFK-AVG basis set also has more accurate mass fractions in the Y (0.8 -- 5~Myr) age bin for the 3 other burst star formation histories. As a result, the DFK-AVG basis set more accurately resolves the burst for the two peaking star formation histories as seen in the right panel of Fig. \ref{fig:rcfig}.

Even though the TB and DFK-AVG basis sets mean mass fractions do not paint a clear picture of which basis set may be preferred overall, the mean light fractions in Fig. \ref{fig:rcfig} for the DFK-AVG basis set show some promise. Looking at the trends in the mean light fractions, we can see that both basis sets can distinguish a galaxy undergoing a burst now (middle two panels) and a galaxy in a post-starburst phase (bottom two panels). Where the DFK-AVG basis set shines, however, is in correctly predicting the relative light fractions for the bursting galaxy. The means for the TB light fractions in the middle two panels incorrectly predict the age of the stellar populations producing more light. The capability of accurately distinguishing a bursting and post-starburst galaxy star formation history could make using the DFK-AVG basis set a useful tool in observational tests of galaxy evolution models.

From Fig. \ref{fig:rcfig} another important drawback of the TB basis set can be illustrated. Looking at the results for the early type galaxy observed at the peak of a burst of star formation (third panel on the left), one can see that the TB basis set recovers an inaccurate mean mass fraction  in the I1 age bin--more than 100 per cent higher. However, looking at the adjacent light fraction panel, TB seems to recover a mean light fraction only about 50 per cent different than the input in that very age bin. The source of this apparent mismatch is the TB basis set's use of SSPs that can match the light of an input spectrum but can have disparate mass-to-light ratios. The I1 age bin contains 5 TB SSPs. The TB basis set in modeling can choose any of these spectra to match the light from stars in this age bin, but the mass-to-light ratio of the oldest SSP in this age bin is nearly 8 times larger than the mass-to-light ratio of the youngest SSP in this bin. Consequently, unless the appropriate mix of SSPs is chosen, the TB basis set can fail to recover an accurate mass fraction while performing relatively better at recovering light fractions.

In summary, we find both the DFK-AVG and TB basis sets were able to accurately recover galaxy $A_{V}$ for galaxies undergoing short bursts.  We found that the DFK-AVG basis set recovered the most precise parameter values. Additionally, we found in general the DFK-AVG basis set recovered more accurate mean mass fractions in the youngest age bin. The TB basis set generally found more accurate mass fractions in the intermediate age bins. Given these results, it is difficult to endorse one basis set over the other in analyzing the star formation histories of quasar host galaxies. Thus, we test both basis sets by analyzing synthetic quasar host galaxy data and a published quasar host galaxy.

\begin{figure*}
\includegraphics{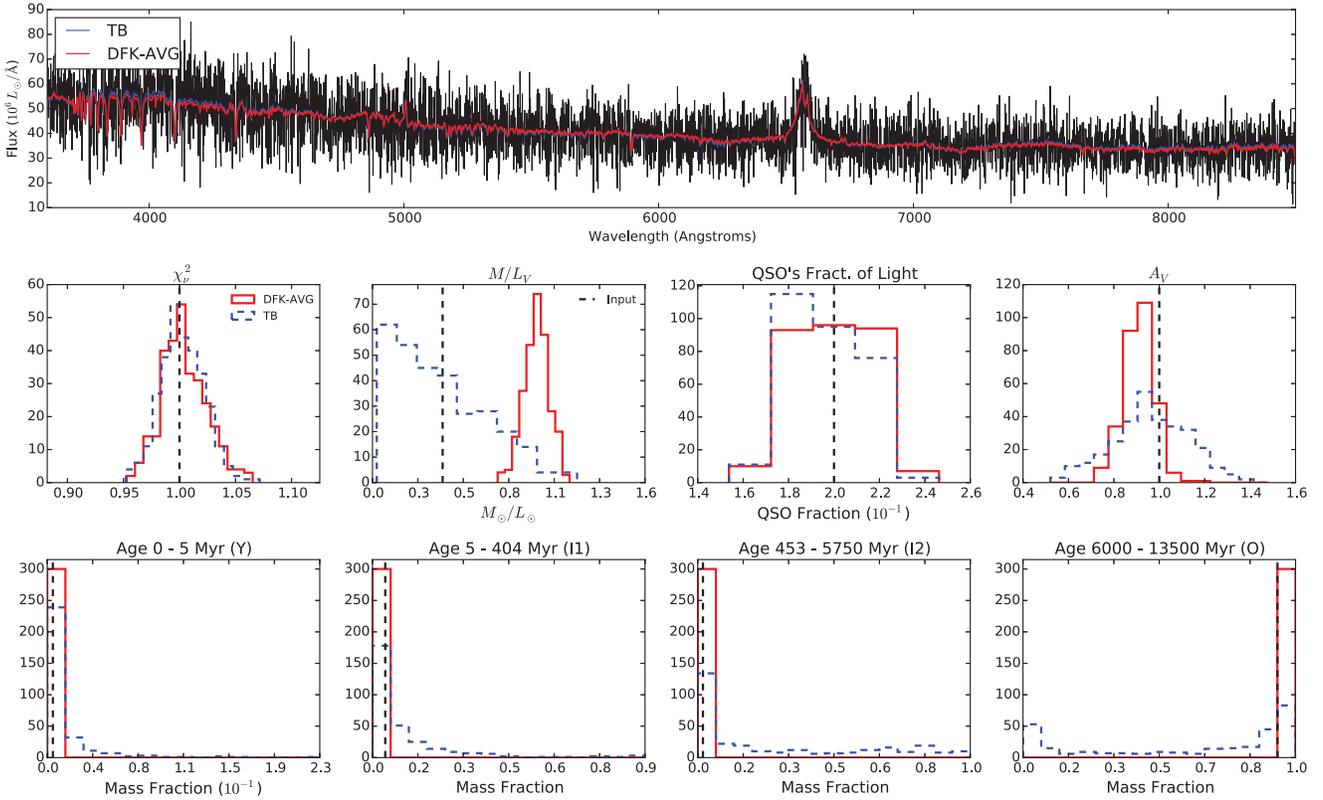}
\caption{The results of running \textsc{sspmodel} on 300 realizations of a synthetic quasar host galaxy spectrum (S/N $\sim$ 5~\AA$^{-1}$) with an underlying early-type ($\tau=$ 2~Gyr) galaxy undergoing a Gaussian burst (observed at the burst peak) for the DFK-AVG and TB basis sets. The top panel shows one of the input spectra in black with a best-fitting spectra for each of the two basis sets over-plotted in color. In the middle and lower panels the results of the DFK-AVG basis are shown as a red solid line and TB as a blue dashed line. The middle panel shows histograms of the reduced $\chi^2$, the \textit{V}-band mass-to-light ratio,  the fraction of light in the spectrum (summed over $\lambda=3600-8500$~\AA) that is scattered QSO light, and the \textit{V}-band dust attenuation. The bottom row shows the histograms of the recovered mass fractions in each age bin. Note that in all histograms the horizontal axis was chosen to display all values selected by each basis set. The input or expected values are plotted as vertical dashed black lines in the histogram plots. The DFK-AVG basis set again produces more precise and accurate results than the TB basis set. The long tails on the TB histograms suggest the larger basis set of SSPs has trouble distinguishing the stellar populations present.}
\label{fig:tau2qhg}
\end{figure*}

\begin{figure*}
\includegraphics{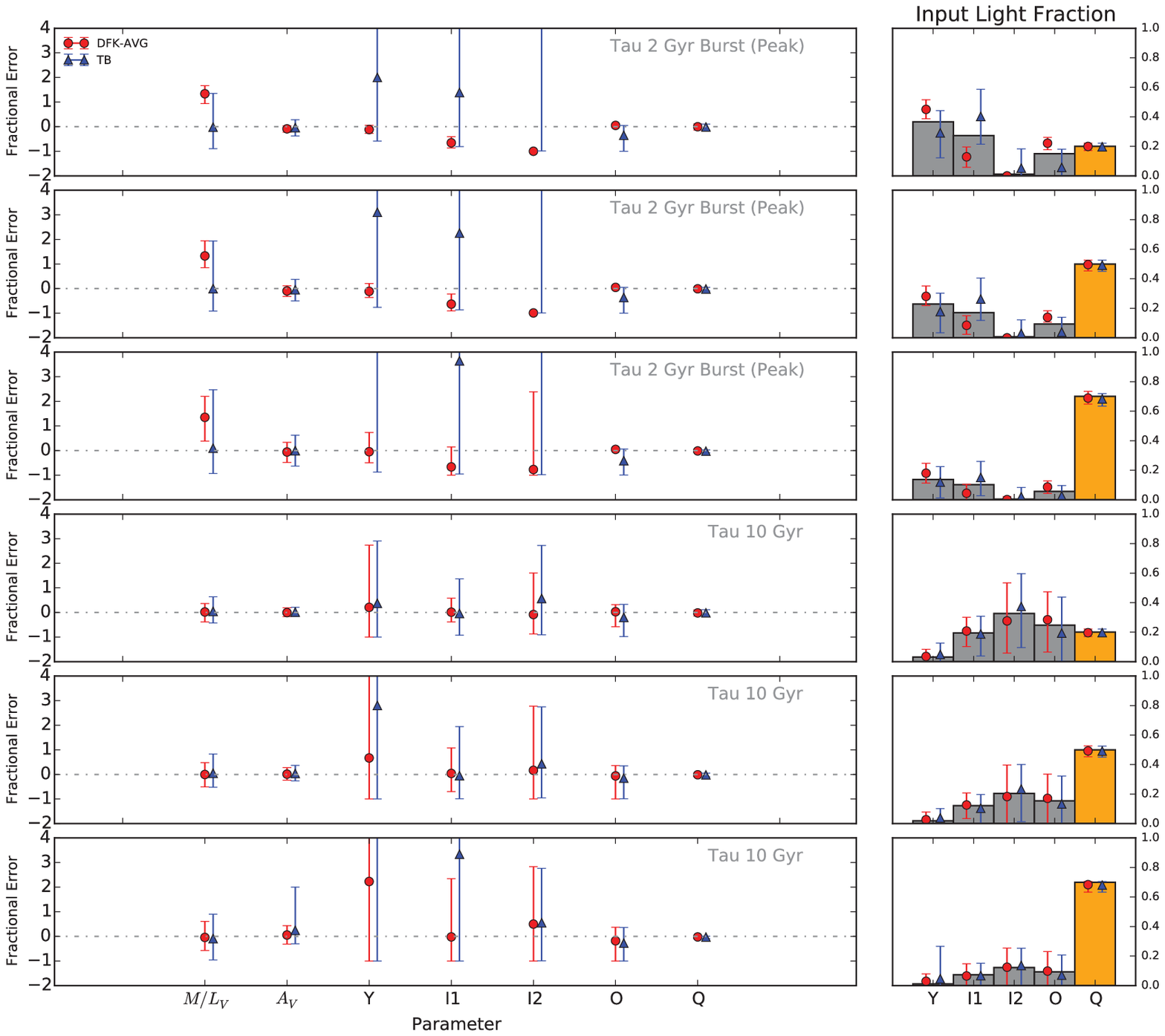}
\caption{The results of \textsc{sspmodel} for the DFK-AVG and TB basis sets for the 2 star formation histories tested with 3 QSO fractions described in \S{\ref{scattqso}}. The fractional error of the mean of $M/L_{V}$, $A_{V}$, and the mass fractions in the 4 broad age bins defined in Table \ref{tab:test-bases-ages} are shown with 95 percentile intervals as error bars in the left series of panels. The error bars show the range of fractional errors that 95 per cent of the recovered values have for each basis set. The input star formation history's light fraction in these broad age bins is shown in the right panels with an additional bar showing the fraction of the light that is QSO scattered light (Q). Again the mean light fraction of each basis set is plotted as points with error bars that show the range of values that 95 per cent of the recovered values have for each basis set. It is apparent in the plot that the DFK-AVG basis set is in general more accurate than the TB basis set--its mean fractional errors lie closer to zero. There are even cases when the TB mean fractional error points do not appear in the plotting region (first 3 rows on the left) as they are even farther away from the input value.}
\label{fig:rcfigqhg}
\end{figure*}

\section{Quasar Host Galaxy Comparison} \label{scattqso}
The goal of this paper is to test a new method of stellar population modeling that could be used on low signal-to-noise spectra, particularly those of quasar host galaxies. In \S\ref{sec:method} and \S\ref{basistesting}, we tested the TB and DFK-AVG basis sets in the simpler scenario of analyzing noisy galaxy spectra without signal from a strong active nucleus. In this section, we put both basis sets to the test on low signal-to-noise longslit observations of simulated quasar host galaxies and the quasar host galaxy PG 0052+251. 

The main difference between modeling quiescent galaxy spectra and modeling quasar host galaxy spectra is that we must now account for and model any quasar light from the wings of the PSF that enter our galaxy observations. This is commonly called scattered quasar light. To model the scattered quasar light present in real quasar host galaxy observations, our code \textsc{sspmodel} accepts as input an on-axis spectrum, acquired with the slit centred on the QSO and an off-axis spectrum, acquired with the slit offset from the QSO. \textsc{sspmodel} then uses an analytic function to estimate the fraction of quasar light scattered by atmospheric turbulence into the off-axis spectrum as a function of wavelength\footnote{The point spread function (PSF) is dependent on wavelength. In general, a PSF dominated by atmospheric seeing gets broader with decreasing wavelength. This produces scattered light in our observations that is more complicated than merely adding a constant multiple of the QSO spectrum.}. Previously, functional approximations (polynomial and power law) were used to model the scattered quasar light \citep{Wold-2010, Miller-2003}. In this paper, however, given a known instrument set-up (slit width, extraction width, and offset of the off-axis spectrum) the only free parameter we need in order to estimate the scattered QSO light is the seeing (see Appendix for details).

\subsection{Model Quasar Host Galaxies} \label{scattqso}
To test whether the TB and DFK-AVG basis sets can recover the stellar populations of quasar host galaxies, we generate 300 noise realizations of quasar host galaxy spectra composed of 20 per cent, 50 per cent, and 70 per cent scattered quasar light for two star formation histories: the exponential decay model, $\tau=10$~Gyr, and the $\tau=2$~Gyr model with a Gaussian burst observed at the burst peak from \S\ref{tau10} and \S\ref{tauburst}, respectively. We use these star formation histories as we might expect recent star formation in quasar host galaxies \citep[e.g.,][]{Dimatteo-2005,Hopkins-2008ii}. We then input the synthetic pairs of observations into \textsc{sspmodel}.

We simulate longslit observations of a quasar host galaxy by generating both on-axis (slit centreed on the QSO) and off-axis (slit centred 2~arcsec from the QSO) observations, modeling the observing technique used in e.g., \citet{Miller-2003}. We use the median SDSS composite quasar spectrum \citep{VandenBerk-2001} uniformly scaled to 6 different luminosities in the \textit{V}-band (see Table \ref{tab:qhgsim}) as our QSO spectrum. These luminosities are chosen to sample 20 per cent, 50 per cent, and 70 per cent scattered QSO light fractions for the two star formation histories tested. These scattered quasar light values refer to the sum total of scattered QSO light divided by total of all light in the off-axis observations. 
\begin{table*}\centering
\caption{The quasar scattered light, seeing, and host galaxy and quasar properties for the synthetic quasar spectra tested in \S{\ref{scattqso}}. The $M_{V}$,  $L_{\mathrm{gal}}$, and $L_{\mathrm{qso}}$ are the total, spatially integrated values.}
\noindent\makebox[\textwidth]{
\begin{tabular}{@{}cccccc@{}}\toprule
SFH &Fraction Scatt. Light & On/Off Seeing (arcsec) & $M_{V}$ (gal)& $M_{V}$ (QSO)& $\log( L_{\mathrm{gal}}/L_{\mathrm{qso}})$ (5100 \AA)\\
\midrule
Tau 2gyr Burst Peak & 0.20 & 1.7 & -24.68 & -26.72 & -0.75\\ 
Tau 2gyr Burst Peak & 0.50 & 1.7 & -24.68 & -28.22 & -1.35 \\
Tau 2gyr Burst Peak & 0.70 & 1.7 & -24.68 & -29.14 & -1.72 \\
Tau 10 Gyr & 0.20 & 1.7 & -23.34 & -25.37 & -0.81 \\
Tau 10 Gyr & 0.50 & 1.7 & -23.34 & -26.87 & -1.41\\
Tau 10 Gyr & 0.70 & 1.7 & -23.34 & -27.79 & -1.78\\
\bottomrule
\end{tabular}
}
\label{tab:qhgsim}
\end{table*}

In forming our synthetic quasar host galaxy observations, we assume a longslit width of 1~arcsec for both the on and off-axis observations. We assume an extraction width (width perpendicular to dispersion direction) of 1~arcsec for the on-axis observation and 3.5~arcsec for the off-axis observation. With these values held fixed, assuming atmospheric seeing dominates the PSF, the fraction of scattered quasar light in the off-axis spectrum is solely determined by the seeing in the on and off-axis observations and the brightness of the QSO. We assume a double Gaussian point spread function (PSF) dominated by the atmospheric seeing parameter in both the on and off-axis observations. A single Gaussian PSF would underestimate the wings of the light profile of the QSO \citep[e.g.,][]{Trujillo-2001}. The double Gaussian avoids this issue. We further make the simplifying assumption that the seeing during the on-axis observation is the same as the seeing for the off-axis observations as typically these observations are done close in time. This means that the fraction of the on-axis observation's light required to match the scattered quasar light in the off-axis spectrum is determined by effectively only one seeing parameter. We fix this seeing parameter at the value 1.7~arcsec for all synthetic observations.

Using the double Gaussian PSF and longslit observation parameters, we calculate the QSO light contained in the extracted slit for both the on and off-axis observations. To form the off-axis observations, we add the calculated QSO light in the off-axis observation to a given galaxy spectrum. We then redden and add noise to this composite (host + scattered QSO) spectrum such that its S/N $\sim$ 5~\AA$^{-1}$. We add noise to the on-axis QSO spectrum such that its S/N $\sim$ 50~\AA$^{-1}$. These two spectra are then used as inputs to \textsc{sspmodel}.

\subsection{Model Quasar Host Galaxy Test Results}
The results of using \textsc{sspmodel} with the DFK-AVG and TB basis sets on the early type galaxy undergoing a burst with 20 per cent scattered QSO light is shown in Fig. \ref{fig:tau2qhg}. The results from all the synthetic host galaxies observations described in \S\ref{scattqso} are shown in Fig. \ref{fig:rcfigqhg}. These results are tabulated as well in Table \ref{tab:qhg-rctable}. The main findings are that for each star formation history and QSO fraction of scattered light, both basis sets were able to accurately recover $A_{V}$. As seen in Fig. \ref{fig:rcfigqhg}, and quantitatively in Table \ref{tab:qhg-rctable}, both TB and DFK-AVG recover $A_{V}$ to within 10 per cent of the input value for each quasar host galaxy scenario tested. However, Table \ref{tab:qhg-rctable} shows that the DFK-AVG basis set recovers more precise values for $A_{V}$ in each quasar host galaxy scenario tested. TB often produces a substantially wider histogram, e.g., as seen in Fig. \ref{fig:tau2qhg}. 

Both basis sets were able to recover an accurate mean amount of scattered quasar light for each quasar host galaxy test. Table \ref{tab:qhg-rctable} shows that each basis set recovers a mean fraction of scattered quasar light ($f_{\mathrm{QSO}}$) to within 3 per cent of the input value. Regarding the star formation history or mean mass fractions recovered, the DFK-AVG basis set recovers more accurate mean mass fractions in every quasar host galaxy scenario tested. As can be seen by studying Fig. \ref{fig:rcfigqhg}, the mean fractional error of the DFK-AVG basis set for the mass fractions in the 4 broad age bins are always closer to zero, i.e. closer to the input value, than the TB basis set. There are even cases when the TB basis set point does not appear in the range of the plot. Quantitatively, in Table \ref{tab:qhg-rctable}, it is clear from comparing the ``Mean Fract'l Error" sub-columns of each basis set for the mass fractions that the DFK-AVG basis set is always closer. Looking at Fig. \ref{fig:rcfigqhg} it is also clear that generally the DFK-AVG basis set is more precise. We also note that the presence of low scattered quasar light (e.g. 20 per cent) only weakly affects the accuracy and precision of the recovered parameters for the DFK-AVG and TB basis sets when compared to tests run without scattered quasar light. In fact, for the early type quasar host galaxy undergoing a burst, we find that the percentage of simulations with recovered values within 25 per cent of the input values only decreases by a small amount for both basis sets when scattered light from a quasar is added and the percentage of values greater than or 100 per cent different than the input does not increase at all for the DFK-AVG results including quasar light. This insensitivity to scattered quasar light, however, may be a function of star formation history. We find that the accuracy of recovered parameters by the DFK-AVG and TB basis sets decreases for the late type quasar host galaxy with low quasar light fraction. However, the DFK-AVG basis set still has more parameter values within 25 per cent of the input values than the TB basis set does.

The panels on the right of Fig. \ref{fig:rcfigqhg} show that both basis sets can distinguish a quasar host galaxy with a more gradual star formation history (bottom 3 panels) from one undergoing a burst (top 3 panels).  However, the DFK-AVG basis set recovers more accurate mean mass and light fractions, in particular for Y and I1 bins of the burst. The TB basis set's mean light fractions for these two bins show an opposite trend similar to the results for the QSO-less synthetic spectra. Accurately distinguishing these two types of star formation histories in quasar host galaxies could make the DFK-AVG basis a useful tool in exploring the relationship between QSO activity and the star formation of the host galaxy.

The accuracy and the precision of the DFK-AVG basis set suggest that the DFK-AVG basis set is the favored choice if one wants to recover star formation histories from low signal-to-noise quasar host galaxies.

\begin{table*}\centering
\caption{Table of results from using the DFK-AVG and TB basis sets to analyze synthetic galaxy spectra for 2 star formation histories at a variety of scattered quasar light percentages discussed in \S{\ref{scattqso}}. The layout is the same as in Table \ref{tab:rctable}. A careful comparison of the mean fractional errors of both basis sets reveals that on average the DFK-AVG basis set performs better than the TB basis set. These results are visually represented in Figure \ref{fig:rcfigqhg}.}
\resizebox{\linewidth}{!}{
\begin{tabular}{@{}ccccccccccc@{}}\toprule
&  \multicolumn{2}{c}{Input}  &  \multicolumn{4}{c}{DFK-AVG} &  \multicolumn{4}{c}{TB}\\
\cmidrule(l{30pt}r{30pt}){2-3}
\cmidrule(l{5pt}r{5pt}){4-7}
\cmidrule(l{5pt}r{5pt}){8-11}
\multicolumn{3}{c}{} & Mean & Median & Mean Frac'l Error & 95 Percentile & Mean & Median & Mean Frac'l Error & 95 Percentile \\
\cmidrule(l{5pt}r{5pt}){4-4}
\cmidrule(l{5pt}r{5pt}){5-5}
\cmidrule(l{5pt}r{5pt}){6-6}
\cmidrule(l{5pt}r{5pt}){7-7}
\cmidrule(l{5pt}r{5pt}){8-8}
\cmidrule(l{5pt}r{5pt}){9-9}
\cmidrule(l{5pt}r{5pt}){10-10}
\cmidrule(l{5pt}r{5pt}){11-11}
& $A_V$ & 1.000 & 0.912 & 0.912 & \textbf{-0.088} & \textbf{(-0.226, 0.035)}  & 0.959 & 0.957 & \textbf{-0.041} & \textbf{(-0.380, 0.283)} \\
\multicolumn{1}{c}{Tau 2 Gyr Burst (Peak)} & $M/L_{V}$ & 0.412 & 0.963 & 0.968 & \textbf{1.338} & \textbf{(0.937, 1.666)}  & 0.401 & 0.350 & \textbf{-0.025} & \textbf{(-0.892, 1.351)} \\
& $f_{QSO}$ & 0.200 & 0.199 & 0.200 & \textbf{-0.003} & \textbf{(-0.143, 0.110)}  & 0.197 & 0.199 & \textbf{-0.014} & \textbf{(-0.142, 0.106)} \\
& $\chi_{\nu}^{2}$ & 1.000 & 1.004 & 1.001 & \textbf{0.004} & \textbf{(-0.032, 0.048)}  & 1.003 & 1.003 & \textbf{0.003} & \textbf{(-0.033, 0.042)} \\
\\
& Light Fraction & Mass Fraction & \\
\cmidrule(l{5pt}r{5pt}){2-2}
\cmidrule(l{5pt}r{5pt}){3-3}
\multicolumn{1}{r}{Y} & 0.458 & 0.005 & 0.004 & 0.004 & \textbf{-0.117} & \textbf{(-0.275, 0.055)}  & 0.014 & 0.006 & \textbf{1.998} & \textbf{(-0.590, 15.333)} \\
\multicolumn{1}{r}{I1} & 0.341 & 0.043 & 0.015 & 0.014 & \textbf{-0.656} & \textbf{(-0.865, -0.405)}  & 0.103 & 0.050 & \textbf{1.387} & \textbf{(-0.813, 11.378)} \\
\multicolumn{1}{r}{I2} & 0.014 & 0.018 & 0.000 & 0.000 & \textbf{-1.000} & \textbf{(-1.000, -1.000)}  & 0.277 & 0.102 & \textbf{14.038} & \textbf{(-0.989, 48.437)} \\
\multicolumn{1}{r}{O} & 0.188 & 0.934 & 0.981 & 0.982 & \textbf{0.051} & \textbf{(0.038, 0.060)}  & 0.606 & 0.785 & \textbf{-0.351} & \textbf{(-0.999, 0.051)} \\
\midrule
& $A_V$ & 1.000 & 0.906 & 0.906 & \textbf{-0.094} & \textbf{(-0.317, 0.114)}  & 0.953 & 0.938 & \textbf{-0.047} & \textbf{(-0.504, 0.375)} \\
\multicolumn{1}{c}{Tau 2 Gyr Burst (Peak)} & $M/L_{V}$ & 0.412 & 0.958 & 0.951 & \textbf{1.328} & \textbf{(0.850, 1.940)}  & 0.411 & 0.278 & \textbf{-0.003} & \textbf{(-0.914, 1.938)} \\
& $f_{QSO}$ & 0.500 & 0.496 & 0.500 & \textbf{-0.008} & \textbf{(-0.095, 0.053)}  & 0.492 & 0.499 & \textbf{-0.017} & \textbf{(-0.098, 0.053)} \\
& $\chi_{\nu}^{2}$ & 1.000 & 1.003 & 1.003 & \textbf{0.003} & \textbf{(-0.031, 0.040)}  & 1.004 & 1.004 & \textbf{0.004} & \textbf{(-0.038, 0.044)} \\
\\
& Light Fraction & Mass Fraction & \\
\cmidrule(l{5pt}r{5pt}){2-2}
\cmidrule(l{5pt}r{5pt}){3-3}
\multicolumn{1}{r}{Y} & 0.458 & 0.005 & 0.004 & 0.004 & \textbf{-0.112} & \textbf{(-0.361, 0.206)}  & 0.019 & 0.007 & \textbf{3.102} & \textbf{(-0.757, 27.459)} \\
\multicolumn{1}{r}{I1} & 0.341 & 0.043 & 0.016 & 0.015 & \textbf{-0.630} & \textbf{(-0.909, -0.218)}  & 0.140 & 0.062 & \textbf{2.254} & \textbf{(-0.865, 18.404)} \\
\multicolumn{1}{r}{I2} & 0.014 & 0.018 & 0.000 & 0.000 & \textbf{-0.989} & \textbf{(-1.000, -0.999)}  & 0.246 & 0.094 & \textbf{12.352} & \textbf{(-0.990, 46.297)} \\
\multicolumn{1}{r}{O} & 0.188 & 0.934 & 0.980 & 0.981 & \textbf{0.049} & \textbf{(0.030, 0.062)}  & 0.594 & 0.763 & \textbf{-0.364} & \textbf{(-0.998, 0.055)} \\
\midrule
& $A_V$ & 1.000 & 0.941 & 0.928 & \textbf{-0.059} & \textbf{(-0.482, 0.336)}  & 0.993 & 0.983 & \textbf{-0.007} & \textbf{(-0.631, 0.624)} \\
\multicolumn{1}{c}{Tau 2 Gyr Burst (Peak)} & $M/L_{V}$ & 0.412 & 0.968 & 0.976 & \textbf{1.350} & \textbf{(0.380, 2.205)}  & 0.450 & 0.282 & \textbf{0.093} & \textbf{(-0.934, 2.470)} \\
& $f_{QSO}$ & 0.700 & 0.689 & 0.698 & \textbf{-0.016} & \textbf{(-0.073, 0.050)}  & 0.682 & 0.696 & \textbf{-0.025} & \textbf{(-0.095, 0.027)} \\
& $\chi_{\nu}^{2}$ & 1.000 & 1.004 & 1.004 & \textbf{0.004} & \textbf{(-0.033, 0.048)}  & 1.008 & 1.008 & \textbf{0.008} & \textbf{(-0.035, 0.052)} \\
\\
& Light Fraction & Mass Fraction & \\
\cmidrule(l{5pt}r{5pt}){2-2}
\cmidrule(l{5pt}r{5pt}){3-3}
\multicolumn{1}{r}{Y} & 0.458 & 0.005 & 0.004 & 0.004 & \textbf{-0.048} & \textbf{(-0.500, 0.735)}  & 0.033 & 0.007 & \textbf{5.952} & \textbf{(-0.879, 40.029)} \\
\multicolumn{1}{r}{I1} & 0.341 & 0.043 & 0.014 & 0.012 & \textbf{-0.665} & \textbf{(-1.000, 0.146)}  & 0.200 & 0.056 & \textbf{3.645} & \textbf{(-0.956, 20.776)} \\
\multicolumn{1}{r}{I2} & 0.014 & 0.018 & 0.004 & 0.000 & \textbf{-0.771} & \textbf{(-1.000, 2.383)}  & 0.219 & 0.082 & \textbf{10.876} & \textbf{(-0.980, 46.513)} \\
\multicolumn{1}{r}{O} & 0.188 & 0.934 & 0.977 & 0.983 & \textbf{0.046} & \textbf{(-0.009, 0.067)}  & 0.548 & 0.678 & \textbf{-0.413} & \textbf{(-0.996, 0.061)} \\
\midrule
& $A_V$ & 1.000 & 0.996 & 0.994 & \textbf{-0.004} & \textbf{(-0.168, 0.193)}  & 1.014 & 1.012 & \textbf{0.014} & \textbf{(-0.158, 0.211)} \\
\multicolumn{1}{c}{Tau 10 Gyr} & $M/L_{V}$ & 1.397 & 1.426 & 1.443 & \textbf{0.021} & \textbf{(-0.384, 0.363)}  & 1.459 & 1.445 & \textbf{0.045} & \textbf{(-0.428, 0.639)} \\
& $f_{QSO}$ & 0.200 & 0.196 & 0.199 & \textbf{-0.018} & \textbf{(-0.147, 0.105)}  & 0.197 & 0.200 & \textbf{-0.016} & \textbf{(-0.146, 0.110)} \\
& $\chi_{\nu}^{2}$ & 1.000 & 1.000 & 1.001 & \textbf{0.000} & \textbf{(-0.037, 0.035)}  & 1.002 & 1.002 & \textbf{0.002} & \textbf{(-0.039, 0.044)} \\
\\
& Light Fraction & Mass Fraction & \\
\cmidrule(l{5pt}r{5pt}){2-2}
\cmidrule(l{5pt}r{5pt}){3-3}
\multicolumn{1}{r}{Y} & 0.039 & 0.0 & 0.000 & 0.000 & \textbf{0.209} & \textbf{(-1.000, 2.738)}  & 0.000 & 0.000 & \textbf{0.373} & \textbf{(-0.999, 2.903)} \\
\multicolumn{1}{r}{I1} & 0.244 & 0.014 & 0.015 & 0.014 & \textbf{0.013} & \textbf{(-0.386, 0.584)}  & 0.014 & 0.013 & \textbf{-0.040} & \textbf{(-0.922, 1.368)} \\
\multicolumn{1}{r}{I2} & 0.408 & 0.26 & 0.238 & 0.198 & \textbf{-0.084} & \textbf{(-0.872, 1.604)}  & 0.407 & 0.321 & \textbf{0.565} & \textbf{(-0.905, 2.721)} \\
\multicolumn{1}{r}{O} & 0.309 & 0.725 & 0.747 & 0.787 & \textbf{0.030} & \textbf{(-0.580, 0.313)}  & 0.578 & 0.665 & \textbf{-0.202} & \textbf{(-0.982, 0.331)} \\
\midrule
& $A_V$ & 1.000 & 1.011 & 1.002 & \textbf{0.011} & \textbf{(-0.236, 0.280)}  & 1.037 & 1.034 & \textbf{0.037} & \textbf{(-0.260, 0.370)} \\
\multicolumn{1}{c}{Tau 10 Gyr} & $M/L_{V}$ & 1.397 & 1.393 & 1.439 & \textbf{-0.003} & \textbf{(-0.505, 0.479)}  & 1.464 & 1.409 & \textbf{0.048} & \textbf{(-0.518, 0.830)} \\
& $f_{QSO}$ & 0.500 & 0.493 & 0.499 & \textbf{-0.015} & \textbf{(-0.095, 0.053)}  & 0.491 & 0.498 & \textbf{-0.018} & \textbf{(-0.099, 0.051)} \\
& $\chi_{\nu}^{2}$ & 1.000 & 1.004 & 1.005 & \textbf{0.004} & \textbf{(-0.038, 0.045)}  & 1.003 & 1.004 & \textbf{0.003} & \textbf{(-0.036, 0.043)} \\
\\
& Light Fraction & Mass Fraction & \\
\cmidrule(l{5pt}r{5pt}){2-2}
\cmidrule(l{5pt}r{5pt}){3-3}
\multicolumn{1}{r}{Y} & 0.039 & 0.0 & 0.000 & 0.000 & \textbf{0.669} & \textbf{(-1.000, 5.535)}  & 0.001 & 0.000 & \textbf{2.803} & \textbf{(-0.999, 5.437)} \\
\multicolumn{1}{r}{I1} & 0.244 & 0.014 & 0.015 & 0.014 & \textbf{0.041} & \textbf{(-0.698, 1.072)}  & 0.014 & 0.009 & \textbf{-0.053} & \textbf{(-0.988, 1.943)} \\
\multicolumn{1}{r}{I2} & 0.408 & 0.26 & 0.303 & 0.200 & \textbf{0.166} & \textbf{(-1.000, 2.773)}  & 0.372 & 0.268 & \textbf{0.429} & \textbf{(-0.957, 2.746)} \\
\multicolumn{1}{r}{O} & 0.309 & 0.725 & 0.681 & 0.785 & \textbf{-0.061} & \textbf{(-1.000, 0.362)}  & 0.614 & 0.717 & \textbf{-0.154} & \textbf{(-0.992, 0.349)} \\
\midrule
& $A_V$ & 1.000 & 1.055 & 1.048 & \textbf{0.055} & \textbf{(-0.317, 0.436)}  & 1.242 & 1.089 & \textbf{0.242} & \textbf{(-0.298, 2.001)} \\
\multicolumn{1}{c}{Tau 10 Gyr} & $M/L_{V}$ & 1.397 & 1.337 & 1.284 & \textbf{-0.043} & \textbf{(-0.575, 0.610)}  & 1.269 & 1.239 & \textbf{-0.091} & \textbf{(-0.962, 0.906)} \\
& $f_{QSO}$ & 0.700 & 0.684 & 0.697 & \textbf{-0.024} & \textbf{(-0.096, 0.007)}  & 0.680 & 0.668 & \textbf{-0.028} & \textbf{(-0.096, 0.006)} \\
& $\chi_{\nu}^{2}$ & 1.000 & 1.006 & 1.005 & \textbf{0.006} & \textbf{(-0.033, 0.049)}  & 1.008 & 1.008 & \textbf{0.008} & \textbf{(-0.027, 0.049)} \\
\\
& Light Fraction & Mass Fraction & \\
\cmidrule(l{5pt}r{5pt}){2-2}
\cmidrule(l{5pt}r{5pt}){3-3}
\multicolumn{1}{r}{Y} & 0.039 & 0.0 & 0.001 & 0.000 & \textbf{2.230} & \textbf{(-1.000, 10.475)}  & 0.009 & 0.000 & \textbf{49.779} & \textbf{(-0.999, 700.875)} \\
\multicolumn{1}{r}{I1} & 0.244 & 0.014 & 0.014 & 0.012 & \textbf{-0.022} & \textbf{(-1.000, 2.342)}  & 0.063 & 0.010 & \textbf{3.335} & \textbf{(-0.998, 56.901)} \\
\multicolumn{1}{r}{I2} & 0.408 & 0.26 & 0.390 & 0.288 & \textbf{0.499} & \textbf{(-1.000, 2.834)}  & 0.404 & 0.327 & \textbf{0.552} & \textbf{(-0.993, 2.767)} \\
\multicolumn{1}{r}{O} & 0.309 & 0.725 & 0.595 & 0.698 & \textbf{-0.179} & \textbf{(-1.000, 0.370)}  & 0.524 & 0.601 & \textbf{-0.278} & \textbf{(-0.999, 0.363)} \\
\bottomrule
\end{tabular}
}

\label{tab:qhg-rctable}
\end{table*}

\subsection{PG0052+251 Test Results}
Overall the DFK-AVG basis set found more precise and accurate mean values of galaxy properties for the synthetic quasar host galaxies tested. But to show that these results can be extended to real observations, we run multiple realizations of degraded Keck LRIS observations of the galaxy PG0052+251 \citep{Sheinis-2002,Wold-2010} through our code with both the TB and DFK-AVG basis sets. These data are longslit observations of both the QSO and the host galaxy at a position offset from the QSO by 3~arcsec. These observations were selected because they were taken at low airmass so that effects from atmospheric refraction could be minimized since these data were taken in 1997 before Keck had an atmospheric dispersion corrector. The signal-to-noise of these data are higher than the synthetic data we tested in earlier sections (S/N $\sim$ 23~\AA$^{-1}$), so we can use the results of \textsc{sspmodel} with these higher signal-to-noise data as a reference and see how recovered galaxy properties might change with lower signal-to-noise. 

The results of running \textsc{sspmodel} on the high signal-to-noise data with the DFK-AVG basis set were consistent with previous studies of this galaxy. PG 0052+251 is classified as a spiral galaxy \citep{Hamilton-2002,Bahcall-1997} and we find using the DFK-AVG basis set that the dominant stellar populations are from the I1 and I2 age bins--indicative of younger stellar populations. 

We can also calculate the light weighted logarithmic age, $\langle\log{t}\rangle_{L}$, of this object's stellar populations as done by \citet{Wold-2010}--their equation 3--in order to compare results. For consistency we use the TB basis set and the same masking. \citeauthor{Wold-2010} find $\langle\log{t}\rangle_{L} = 8.18 \pm 0.34$ whereas we find $\langle\log{t}\rangle_{L} = 8.64 \pm 0.14$. Monte Carlo simulations confirm that much of the difference in both the mean and the error bars is due to the adopted method of modeling the quasar scattered light.  Notably, the error bars decrease by a factor 2.4  when switching from \citeauthor{Wold-2010}'s method of using a polynomial with four free parameters to our physically motivated method which uses two free parameters. We find $\langle\log{t}\rangle_{L} = 9.09 \pm 0.08$  using the DFK-AVG basis set. The DFK-AVG $\langle\log{t}\rangle_{L}$ is 3.21$\sigma$ from the TB result. We can use our previous simulations to provide some information in choosing which basis set to trust. Comparing the accuracy of $\langle\log{t}\rangle_{L}$ of the two basis sets across all 12 star formation and quasar host galaxy scenarios tested, we find that the DFK-AVG basis set's estimates are closer to the input $\langle\log{t}\rangle_{L}$ 75 per cent of the time (9/12). Therefore, we adopt the DFK-AVG  $\langle\log{t}\rangle_{L}$ as the more accurate value.

To degrade the higher signal-to-noise data of PG 0052+251, we add Gaussian noise to the offset host galaxy observation so that the median S/N $\sim$ 5~\AA$^{-1}$. We do not add any noise to the QSO, on-axis, observation. We make 300 realizations of these degraded observations and run them through \textsc{sspmodel} for both the TB and DFK-AVG basis sets.

\begin{figure*}
\includegraphics{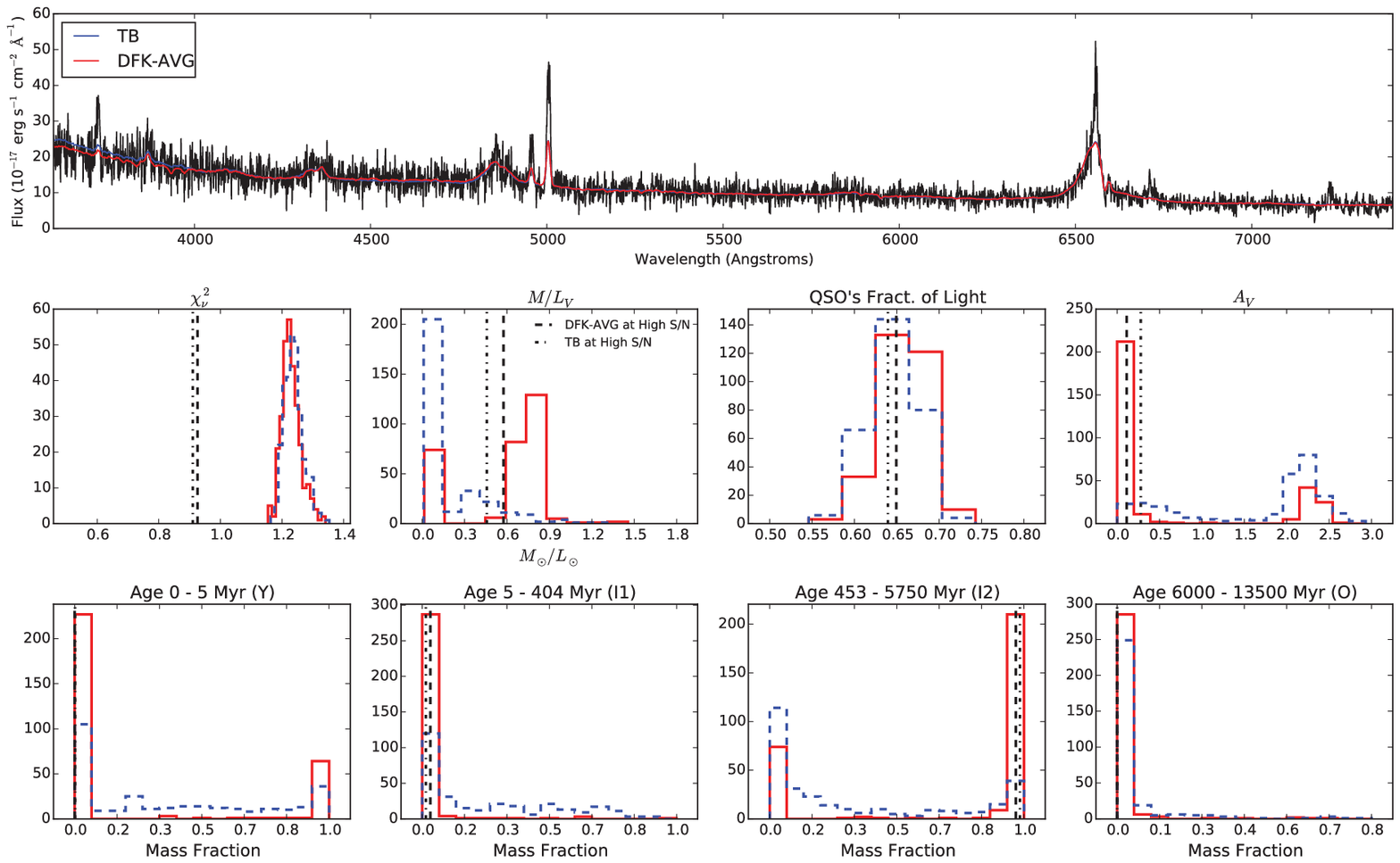}
\caption{The results of running \textsc{sspmodel} on 300 realizations of a PG 0052+251 with S/N$\sim$5~\AA$^{-1}$ for the DFK-AVG and TB basis sets. See Fig. \ref{fig:tau2qhg} for plot description. In this plot, the dashed vertical line represents the results of \textsc{sspmodel} using the DFK-AVG basis set on the higher signal-to-noise (not degraded) spectrum of PG 0052+251. The dot-dashed line represents the results from using the TB basis set on the higher signal-to-noise spectrum. Note that both basis sets reproduce the input noisy spectrum, and though both basis sets have some difficulty recovering galaxy parameters at this S/N, the DFK-AVG basis set consistently recovers more precise parameters consistent with the higher signal-to-noise results; it has fewer misses, suggesting the DFK-AVG basis set is less sensitive to noise.}
\label{fig:pgfig}
\end{figure*}

The results of running PG0052+251's degraded observations with the two basis sets are summarized in Fig. \ref{fig:pgfig}, a histogram plot similar to the previous plots but with the results from the original un-modified higher signal-to-noise observations shown as vertical lines. Fig. \ref{fig:pgfig} shows both basis sets fit the input noisier spectrum of PG 0052+251. The two basis sets are only distinguishable in the bluer wavelengths of the spectrum for the sample fits shown. The basis sets' distribution of  $\chi_{\nu}^{2}$ also have similar means and widths. Both basis sets also have very similar histograms for the recovered QSO fraction demonstrating that both basis sets have similar ease in recovering this parameter at this signal-to-noise.

However, though both basis sets recover QSO fractions that are consistent with one another and the higher signal-to-noise data, it is clear in Fig. \ref{fig:pgfig} that the TB basis set is much less precise in recovering the host galaxy attenuation, $A_{V}$, and the Y, I1, and I2 mass fractions. The longer and more prominent tails in the TB results  in the Y, I1 and I2 histograms suggest that once again the TB basis set has trouble attributing the galaxy's light to these age bins, some times attributing the wrong mixture. The DFK-AVG basis set is not without its misses as well in recovering $A_{V}$ and the Y and I2 mass fractions. But while the TB histograms for these parameters either lack a strong peak near the higher signal-to-noise result or even get an inconsistent peak, the DFK-AVG basis set consistently has more values closer to the higher signal-to-noise results.

In summary, we find satisfactory agreement between our results from runs of the high signal-to-noise data for PG 0052+251 through \textsc{sspmodel} and previous studies of the object. For PG 0052+251's degraded data, both basis sets are able to fit the noisier input spectrum and recover the QSO fraction consistent with the results from analyzing the higher signal-to-noise data. But, the DFK-AVG basis set was more precise and had results that deviated less from the higher signal-to-noise result, especially for the galaxy attenuation, $A_{V}$, and star formation history (Y, I1, and I2 mass fractions).

\begin{table*}\centering
\caption{The rankings by parameter of interest of each basis set tested for the early type and late type galaxy spectra simulations. Under each basis set heading there a three columns representing 3 metrics used to judge each basis set: (1) the mean fraction of the recovered values at and within 25 per cent of the input value, (2) the mean fraction of recovered values at and within 50 per cent of the input value, and (3) the mean fraction of recovered values at and outside 100 per cent of the input value. The first two metrics measure precision and accuracy. Ranks closer to 1 indicate more recovered values closer to the input value. For example, if a basis set had the highest mean fraction of values within 25 per cent for a parameter, it would receive a rank 1. If the basis set had the lowest mean fraction, it would receive a rank equal to the number of basis sets tested for that star formation history. The last metric measures reliability by quantifying how often a basis set is 100 per cent or more off from the input value. For this metric, if a basis set has the lowest percentage of values 100 per cent or more off from the input, it would receive rank 1. In the case of a tie in comparisons, each basis set receives the same rank it would have had there not been a tie. The DFK-SSP and CONSTLF basis sets do comparably worse than the DFK-AVG basis sets in the first 2 SFHs tested with regard to accuracy, though the CONSTLF basis set seems to be on average just as precise as the DFK-AVG basis set. Both DFK-AVG and CONSTLF do better than TB and DFK-SSP suggesting that averaging SSPs assuming some star formation history is useful and that for noisy data simply lowering the number of SSPs might be less accurate than averaging SSPs.}
\label{tab:sum-metric}
\resizebox{\linewidth}{!}{
\begin{tabular}{@{}ccccccccccccc@{}}\toprule
         &  \multicolumn{4}{c}{Mean $f(|v|\leq 25\%)$ Rank} &  \multicolumn{4}{c}{Mean $f(|v|\leq 50\%)$ Rank} &  \multicolumn{4}{c}{Mean $f(|v|\geq 100\%)$ Rank}\\
\cmidrule(l{5pt}r{5pt}){2-5}
\cmidrule(l{5pt}r{5pt}){6-9}
\cmidrule(l{5pt}r{5pt}){10-13}
\multicolumn{1}{c}{Parameters} & DFK-AVG & TB & DFK-SSP & CONSTLF & DFK-AVG & TB & DFK-SSP & CONSTLF & DFK-AVG & TB & DFK-SSP & CONSTLF \\
\cmidrule(l{5pt}r{5pt}){2-2}
\cmidrule(l{5pt}r{5pt}){3-3}
\cmidrule(l{5pt}r{5pt}){4-4}
\cmidrule(l{5pt}r{5pt}){5-5}
\cmidrule(l{5pt}r{5pt}){6-6}
\cmidrule(l{5pt}r{5pt}){7-7}
\cmidrule(l{5pt}r{5pt}){8-8}
\cmidrule(l{5pt}r{5pt}){9-9}
\cmidrule(l{5pt}r{5pt}){10-10}
\cmidrule(l{5pt}r{5pt}){11-11}
\cmidrule(l{5pt}r{5pt}){12-12}
\cmidrule(l{5pt}r{5pt}){13-13}
$A_V$ & 1  & 2  & 1  & 1  & 1  & 1  & 1  & 1  & 1  & 1  & 1  & 1 \\
$M/L_{{V}}$ & 1  & 2  & 3  & 4  & 1  & 2  & 1  & 1  & 1  & 1  & 1  & 1 \\
Y& 2  & 3  & 4  & 1  & 2  & 3  & 4  & 1  & 3  & 4  & 2  & 1  \\
I1& 1  & 4  & 2  & 3  & 2  & 4  & 1  & 3  & 2  & 4  & 1  & 3  \\
I2& 1  & 2  & 3  & 4  & 1  & 2  & 4  & 3  & 1  & 4  & 3  & 2  \\
O& 2  & 3  & 4  & 1  & 1  & 2  & 3  & 1  & 1  & 1  & 2  & 1  \\
\bottomrule
\end{tabular}
}
\end{table*}
\begin{table*}\centering
\caption{The rankings by parameter of interest for the TB and DFK-AVG basis sets for all 12 synthetic star formation history and quasar host combinations tested (6 star formation histories tested in \S{\ref{testedsfhs}} and the other 6 cases in \S{\ref{scattqso}}). For heading description see Table \ref{tab:sum-metric}. It is clear from the rankings that the DFK-AVG basis set is more accurate than the TB basis set using the mean fraction of values within 25 per cent of the input values. It is also clear that the DFK-AVG basis set is more precise by looking at the third column metric. }
\label{tab:full-sum-metric}
\resizebox{\linewidth}{!}{
\begin{tabular}{@{}ccccccc@{}}\toprule
         &  \multicolumn{2}{c}{Mean $f(|v|\leq 25\%)$ Rank} &  \multicolumn{2}{c}{Mean $f(|v|\leq 50\%)$ Rank} &  \multicolumn{2}{c}{Mean $f(|v|\geq 100\%)$ Rank}\\
\cmidrule(l{5pt}r{5pt}){2-3}
\cmidrule(l{5pt}r{5pt}){4-5}
\cmidrule(l{5pt}r{5pt}){6-7}
\multicolumn{1}{c}{Parameters} & DFK-AVG & TB & DFK-AVG & TB & DFK-AVG & TB \\
\cmidrule(l{5pt}r{5pt}){2-2}
\cmidrule(l{5pt}r{5pt}){3-3}
\cmidrule(l{5pt}r{5pt}){4-4}
\cmidrule(l{5pt}r{5pt}){5-5}
\cmidrule(l{5pt}r{5pt}){6-6}
\cmidrule(l{5pt}r{5pt}){7-7}
$A_V$ & 1  & 2  & 1  & 2  & 1  & 2 \\
$M/L_{{V}}$ & 2  & 1  & 2  & 1  & 2  & 1 \\
Y& 1  & 2  & 1  & 2  & 1  & 2  \\
I1& 1  & 2  & 2  & 1  & 1  & 2  \\
I2& 1  & 2  & 2  & 1  & 1  & 2  \\
O& 1  & 2  & 1  & 2  & 1  & 1  \\
\bottomrule
\end{tabular}
}
\end{table*}

\section{Discussion and Conclusions} \label{disc}
The goal of this work was to develop and test a new spectral fitting by inversion method for recovering star formation histories from low signal-to-noise galaxy spectra with the additional ability to model ``scattered" quasar light. Such a method would be a useful tool in studying quasar host galaxies. The method developed uses diffusion k-means to group SSPs into 4 broad age bins and then averages the SSPs in each age bin to form the DFK-AVG basis set. The results of using this DFK-AVG basis set to analyse 12 unique types of synthetic galaxy spectra were then compared to the results of 3 other basis sets: a traditional basis set (TB)--the 15 solar metalicity SSPs of \citet{Cid-2005}, a basis set of individual SSPs with ages close to the mean ages of the SSPs in the DFK-AVG age bins (DFK-SSP), and a basis set whose bases would contribute equal fractions of the light if star formation was constant for all times (CONSTLF). 

We find it illustrative to use three metrics (fractions) to judge each basis set: (1) the fraction of the recovered values at and within 25 per cent of the input value, (2) the fraction of recovered values at and within 50 per cent of the input value, and (3) the fraction of recovered values at and outside 100 per cent of the input value. The first two metrics measure precision and accuracy. The last metric measures reliability by quantifying how often a basis set is 100 per cent or more off from the input value. We compute these fractions by parameter of interest for each basis set for the star formation history and quasar host combinations tested. We then average these fractions over the star formation histories tested and rank the resultant  means of each basis set for each metric. The ranks of the 4 basis sets over the early-type and late-type star formation histories tested are shown in Table \ref{tab:sum-metric}. The ranks of the TB and DFK-AVG basis sets over all 12 of the star formation history and quasar host combinations tested are shown in Table \ref{tab:full-sum-metric}. A rank of 1 denotes the better performing basis set.

After performing the simpler test of comparing each basis set's results on synthetic galaxies without scattered light from a quasar, we eliminated the DFK-SSP and CONSTLF basis sets for their generally lower level of accuracy in recovering mass fractions in the 4 broad age bins. It is shown in the average rankings of these basis sets in Table \ref{tab:sum-metric} that both basis sets perform more poorly than the DFK-AVG basis set in recovering the mass fractions to within 25 per cent of the input value. The worse performance of the DFK-SSP basis set in comparison to the average DFK-AVG basis set suggests that simply lowering the number of SSPs to model noisy data might not be the best practice. The worse performance of the TB basis set (our control) with respect to the CONSTLF and DFK-AVG basis sets suggest that averaging SSPs with some assumed star formation history is a useful tool. Nevertheless, the lackluster performance of the CONSTLF basis set suggests that physical insight should not be ignored (the CONSTLF has a large O age bin over which the SFH is likely not constant) in deciding which and how many SSPs to average. Additionally, the similarity of the SSP spectra was not taken into account in the groupings of the CONSTLF basis set, which might explain its weaker performance.

The good news, however, is that light weighted age of a galaxy, $\langle\log{t}\rangle_{L}$ is a reliable relative quantity for all basis sets. We can calculate $\langle\log{t}\rangle_{L}$ for our simulation results, and we find that all basis sets can correctly differentiate which galaxies have had more recent star formation (smaller $\langle\log{t}\rangle_{L}$).

\begin{figure*}
\includegraphics{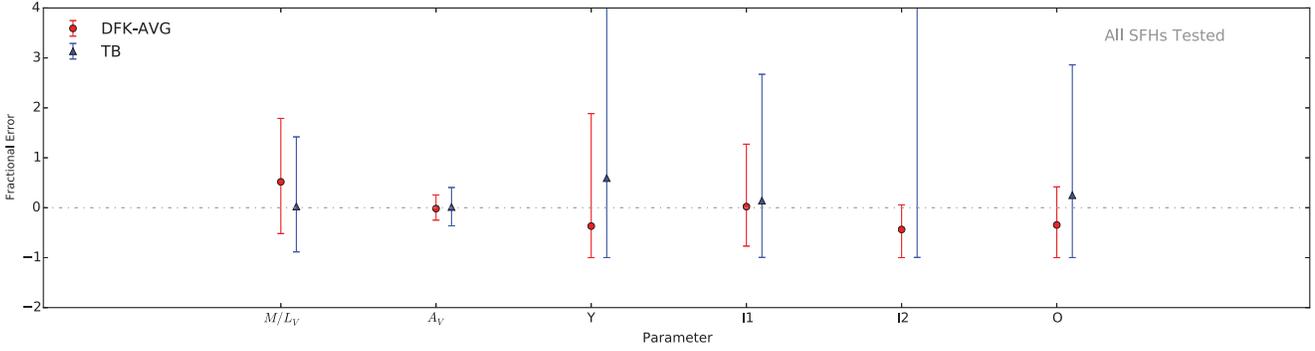}
\caption{The mean fractional error with the 95 percentile interval as error bars for the 6 parameters used to analyze all 12 types of synthetic galaxy spectra tested in this paper for the TB and DFK-AVG basis sets. Looking at the ensemble of results, we see that the DFK-AVG basis set is more precise and often more accurate than the TB basis set.}
\label{fig:rcfullfig}
\end{figure*}

In the comparison of results from the remaining basis sets (DFK-AVG and TB) across synthetic galaxies with and without scattered quasar light, the DFK-AVG basis set was found to be generally more accurate and reliable. As seen in the rankings of Table \ref{tab:full-sum-metric}, DFK-AVG has better ranking in recovering parameters within 25 per cent of the input in all parameters but the mass-to-light ratio, which is a systematic that is easily understood (see \S{\ref{tau2}}). Looking at the last heading of Table \ref{tab:full-sum-metric} comparing the ranks for the DFK-AVG and TB basis sets, it is also clear that the DFK-AVG basis set on average has fewer catastrophic failures, recovering values 100 per cent or more different than the input values. The greater precision and accuracy of the DFK-AVG basis set can be seen even more clearly in a plot of the mean fractional errors of the two basis sets using data from all of the tests on synthetic spectra  in Fig. \ref{fig:rcfullfig}. The shorter extent of the 95 percentile intervals show the increased precision of the DFK-AVG basis set and the smaller mean fractional errors show the accuracy of the basis set.

Our results suggest that for analyzing low signal-to-noise galaxy spectra, the method developed here using diffusion k-means that averages SSPs would be suitable. The DFK-AVG basis set's results were on average more accurate or just as accurate as the TB basis set's in recovering mass fractions in broad age bins, and the DFK-AVG results were clearly shown to be more precise, or reliable.

We showed that our method could be used to analyze the spectrum of a real quasar host galaxy. Moreover, we found that even after introducing artificial noise into the spectrum of the quasar host galaxy data for PG 0052+251, the DFK-AVG basis set could produce results that were more often than not consistent with the higher signal-to-noise results. This suggests that the DFK-AVG basis set is less sensitive to noise than the larger TB basis set which is an attractive property that can have applications that stretch beyond analyzing quasar host galaxy data. As stated previously, depending on the galaxy star formation history, we can predict that there might be a systematic offset in the mass-to-light ratio recovered by the DFK-AVG basis set. However, this offset is small compared to other systematics affecting this parameter (e.g., the IMF).

While the systematics of the DFK-AVG results are a less desirable trait of the basis set, the precision of the basis set is still very valuable. If one analyzes a sample of low signal-to-noise galaxy spectra with identical star formation histories--for example, observed undergoing a burst--our results in Fig. \ref{fig:rcfig} show that the TB basis set would not recover that this sample of galaxies have similar star formation histories. The DFK-AVG basis set, however, would clearly demonstrate this. Then, in principle, one could co-add/stack the individual spectra that have similar star formation histories to form a spectrum with higher signal-to-noise on which one could use more traditional stellar population modeling techniques as one pleased. So, though there are systematics with the DFK-AVG basis set, its precision alone can still be a useful tool in analyzing a large collection of low signal-to-noise galaxy spectra.

As data from large spectroscopic surveys accumulate, the use of a reduced basis set like the DFK-AVG basis set can become a powerful tool in galaxy evolution studies. We showed in \S{\ref{tauburst}} that the DFK-AVG basis set can distinguish bursting and post-starburst star formation histories. Likewise, we show in the presence of a bright quasar, the DFK-AVG basis set can continue to recover accurate mean mass fractions and relative light fractions. This capability could be useful in probing the relationship between supermassive black holes and their galaxies and testing some of the prevailing ideas regarding the timing of starbursts and black hole activity.\citep[e.g.,][]{Cen-2012,Hopkins-2012,Angles-2013}

This paper outlines one method using diffusion k-means to create a small basis set for stellar population modeling, but the process is not universal. There are several customizable features in the process including: (1) selecting a wavelength range to run diffusion k-means on the SSPs, (2) the tuning parameter, $\epsilon$, that might need to change, and (3) the number of bases in the basis set, $k$, which should be chosen based on the data's signal-to-noise and resolution. The software and packages needed to use diffusion k-means are available publicly through the Comprehensive R Archive Network (CRAN). Interested readers should contact the author for access to \textsc{sspmodel}. The full DFK-AVG basis set spectra will be made available electronically.

\section{Acknowledgments}
This material is based upon work supported by the National Science Foundation under Grant No. DGE-0718123 and the Advanced Opportunity fellowship program at the University of Wisconsin-Madison with funds provided by the State of Wisconsin and from the UW-Madison Graduate School. This research was performed using the compute resources and assistance of the UW-Madison Center For High Throughput Computing (CHTC) in the Department of Computer Sciences. The CHTC is supported by UW-Madison and the Wisconsin Alumni Research Foundation, and is an active member of the Open Science Grid, which is supported by the National Science Foundation and the U.S. Department of Energy's Office of Science. We would also like to thank the very thoughtful referee whose insightful comments helped strengthen this paper. We would also like to thank Dr. Isak Wold who was a valuable resource in the development of this paper. This research has made use of NASA's Astrophysics Data System.

\bibliographystyle{mn2e}
\bibliography{refbackup}

\begin{thebibliography}{60}
\expandafter\ifx\csname natexlab\endcsname\relax\def\natexlab#1{#1}\fi

\bibitem[{{Angl{\'e}s-Alc{\'a}zar}
  {et~al}\mbox{.}(2013){Angl{\'e}s-Alc{\'a}zar}, {{\"O}zel}, {Dav{\'e}},
  {Katz}, {Kollmeier}, \& {Oppenheimer}}]{Angles-2013}
{Angl{\'e}s-Alc{\'a}zar} D., {{\"O}zel} F., {Dav{\'e}} R., {Katz} N.,
  {Kollmeier} J.~A., {Oppenheimer} B.~D., 2013, ArXiv e-prints

\bibitem[{{Bahcall} {et~al}\mbox{.}(1997){Bahcall}, {Kirhakos}, {Saxe}, \&
  {Schneider}}]{Bahcall-1997}
{Bahcall} J.~N., {Kirhakos} S., {Saxe} D.~H., {Schneider} D.~P., 1997, \apj,
  479, 642

\bibitem[{{Brotherton} {et~al}\mbox{.}(1999){Brotherton}, {van Breugel},
  {Stanford}, {Smith}, {Boyle}, {Miller}, {Shanks}, {Croom}, \&
  {Filippenko}}]{Brotherton-1999}
{Brotherton} M.~S. {et~al.}, 1999, \apjl, 520, L87

\bibitem[{{Bruzual} \& {Charlot}(2003)}]{Bruzual-2003}
{Bruzual} G., {Charlot} S., 2003, \mnras, 344, 1000

\bibitem[{{Cales} {et~al}\mbox{.}(2013){Cales}, {Brotherton}, {Shang},
  {Runnoe}, {DiPompeo}, {Bennert}, {Canalizo}, {Hiner}, {Stoll}, {Ganguly}, \&
  {Diamond-Stanic}}]{Cales-2013}
{Cales} S.~L. {et~al.}, 2013, \apj, 762, 90

\bibitem[{{Cano-D{\'{\i}}az} {et~al}\mbox{.}(2012){Cano-D{\'{\i}}az},
  {Maiolino}, {Marconi}, {Netzer}, {Shemmer}, \& {Cresci}}]{Cano-Diaz-2012}
{Cano-D{\'{\i}}az} M., {Maiolino} R., {Marconi} A., {Netzer} H., {Shemmer} O.,
  {Cresci} G., 2012, \aap, 537, L8

\bibitem[{{Cardelli}, {Clayton} \& {Mathis}(1989){Cardelli}, {Clayton}, \&
  {Mathis}}]{Cardelli-1989}
{Cardelli} J.~A., {Clayton} G.~C., {Mathis} J.~S., 1989, \apj, 345, 245

\bibitem[{{Cen}(2012)}]{Cen-2012}
{Cen} R., 2012, \apj, 755, 28

\bibitem[{{Chabrier}(2003)}]{Chabrier-2003}
{Chabrier} G., 2003, \pasp, 115, 763

\bibitem[{{Chen} {et~al}\mbox{.}(2012){Chen}, {Kauffmann}, {Tremonti}, {White},
  {Heckman}, {Kova{\v c}}, {Bundy}, {Chisholm}, {Maraston}, {Schneider},
  {Bolton}, {Weaver}, \& {Brinkmann}}]{Chen-2012}
{Chen} Y.-M. {et~al.}, 2012, \mnras, 421, 314

\bibitem[{{Chilingarian} {et~al}\mbox{.}(2007){Chilingarian}, {Prugniel},
  {Sil'Chenko}, \& {Koleva}}]{Chilingarian-2007}
{Chilingarian} I., {Prugniel} P., {Sil'Chenko} O., {Koleva} M., 2007, in IAU
  Symposium, Vol. 241, IAU Symposium, {Vazdekis} A., {Peletier} R., eds., pp.
  175--176

\bibitem[{{Cid Fernandes} {et~al}\mbox{.}(2004){Cid Fernandes}, {Gonz{\'a}lez
  Delgado}, {Schmitt}, {Storchi-Bergmann}, {Martins}, {P{\'e}rez}, {Heckman},
  {Leitherer}, \& {Schaerer}}]{Cid-2004}
{Cid Fernandes} R. {et~al.}, 2004, \apj, 605, 105

\bibitem[{{Cid Fernandes} {et~al}\mbox{.}(2005){Cid Fernandes}, {Mateus},
  {Sodr{\'e}}, {Stasi{\'n}ska}, \& {Gomes}}]{Cid-2005}
{Cid Fernandes} R., {Mateus} A., {Sodr{\'e}} L., {Stasi{\'n}ska} G., {Gomes}
  J.~M., 2005, \mnras, 358, 363

\bibitem[{{Connolly} {et~al}\mbox{.}(1995){Connolly}, {Szalay}, {Bershady},
  {Kinney}, \& {Calzetti}}]{Connolly-1995}
{Connolly} A.~J., {Szalay} A.~S., {Bershady} M.~A., {Kinney} A.~L., {Calzetti}
  D., 1995, \aj, 110, 1071

\bibitem[{{Conroy} \& {Gunn}(2010)}]{Conroy-2010}
{Conroy} C., {Gunn} J.~E., 2010, \apj, 712, 833

\bibitem[{{Conroy}, {Gunn} \& {White}(2009){Conroy}, {Gunn}, \&
  {White}}]{Conroy-2009}
{Conroy} C., {Gunn} J.~E., {White} M., 2009, \apj, 699, 486

\bibitem[{{Davies} {et~al}\mbox{.}(2007){Davies}, {M{\"u}ller S{\'a}nchez},
  {Genzel}, {Tacconi}, {Hicks}, {Friedrich}, \& {Sternberg}}]{Davies-2007}
{Davies} R.~I., {M{\"u}ller S{\'a}nchez} F., {Genzel} R., {Tacconi} L.~J.,
  {Hicks} E.~K.~S., {Friedrich} S., {Sternberg} A., 2007, \apj, 671, 1388

\bibitem[{{Di Matteo}, {Springel} \& {Hernquist}(2005){Di Matteo}, {Springel},
  \& {Hernquist}}]{Dimatteo-2005}
{Di Matteo} T., {Springel} V., {Hernquist} L., 2005, \nat, 433, 604

\bibitem[{{Hamilton}, {Casertano} \& {Turnshek}(2002){Hamilton}, {Casertano},
  \& {Turnshek}}]{Hamilton-2002}
{Hamilton} T.~S., {Casertano} S., {Turnshek} D.~A., 2002, \apj, 576, 61

\bibitem[{{Heavens}, {Jimenez} \& {Lahav}(2000){Heavens}, {Jimenez}, \&
  {Lahav}}]{Heavens-2000}
{Heavens} A.~F., {Jimenez} R., {Lahav} O., 2000, \mnras, 317, 965

\bibitem[{{Hopkins}(2012)}]{Hopkins-2012}
{Hopkins} P.~F., 2012, \mnras, 420, L8

\bibitem[{{Hopkins} {et~al}\mbox{.}(2008){Hopkins}, {Cox}, {Kere{\v s}}, \&
  {Hernquist}}]{Hopkins-2008ii}
{Hopkins} P.~F., {Cox} T.~J., {Kere{\v s}} D., {Hernquist} L., 2008, \apjs,
  175, 390

\bibitem[{{Hopkins} {et~al}\mbox{.}(2006){Hopkins}, {Hernquist}, {Cox}, {Di
  Matteo}, {Robertson}, \& {Springel}}]{Hopkins-2006}
{Hopkins} P.~F., {Hernquist} L., {Cox} T.~J., {Di Matteo} T., {Robertson} B.,
  {Springel} V., 2006, \apjs, 163, 1

\bibitem[{{Jahnke} {et~al}\mbox{.}(2004){Jahnke}, {Wisotzki}, {S{\'a}nchez},
  {Christensen}, {Becker}, {Kelz}, \& {Roth}}]{Jahnke-2004b}
{Jahnke} K., {Wisotzki} L., {S{\'a}nchez} S.~F., {Christensen} L., {Becker} T.,
  {Kelz} A., {Roth} M.~M., 2004, Astronomische Nachrichten, 325, 128

\bibitem[{{Johansson}, {Thomas} \& {Maraston}(2012){Johansson}, {Thomas}, \&
  {Maraston}}]{Johansson-2012}
{Johansson} J., {Thomas} D., {Maraston} C., 2012, \mnras, 421, 1908

\bibitem[{{Koleva} {et~al}\mbox{.}(2009){Koleva}, {Prugniel}, {Bouchard}, \&
  {Wu}}]{Koleva-2009}
{Koleva} M., {Prugniel} P., {Bouchard} A., {Wu} Y., 2009, \aap, 501, 1269

\bibitem[{{Lafon} \& {Lee}(2006)}]{Lafon-2006}
{Lafon} S., {Lee} A.~B., 2006, IEEE Transactions on Pattern Analysis and
  Machine Intelligence, 28, 1393

\bibitem[{{Le Borgne} {et~al}\mbox{.}(2003){Le Borgne}, {Bruzual}, {Pell{\'o}},
  {Lan{\c c}on}, {Rocca-Volmerange}, {Sanahuja}, {Schaerer}, {Soubiran}, \&
  {V{\'{\i}}lchez-G{\'o}mez}}]{Leborgne-2003}
{Le Borgne} J.-F. {et~al.}, 2003, \aap, 402, 433

\bibitem[{{Lu} {et~al}\mbox{.}(2006){Lu}, {Zhou}, {Wang}, {Wang}, {Dong},
  {Zhuang}, \& {Li}}]{Lu-2006}
{Lu} H., {Zhou} H., {Wang} J., {Wang} T., {Dong} X., {Zhuang} Z., {Li} C.,
  2006, \aj, 131, 790

\bibitem[{{Madgwick} {et~al}\mbox{.}(2003){Madgwick}, {Somerville}, {Lahav}, \&
  {Ellis}}]{Madgwick-2003}
{Madgwick} D.~S., {Somerville} R., {Lahav} O., {Ellis} R., 2003, \mnras, 343,
  871

\bibitem[{{Mathis}, {Charlot} \& {Brinchmann}(2006){Mathis}, {Charlot}, \&
  {Brinchmann}}]{Mathis-2006}
{Mathis} H., {Charlot} S., {Brinchmann} J., 2006, \mnras, 365, 385

\bibitem[{{Miller} \& {Sheinis}(2003)}]{Miller-2003}
{Miller} J.~S., {Sheinis} A.~I., 2003, \apjl, 588, L9

\bibitem[{{Murtagh} \& {Heck}(1987)}]{Murtagh-1987}
{Murtagh} F., {Heck} A., eds., 1987, Astrophysics and Space Science Library,
  Vol. 131, {Multivariate Data Analysis}

\bibitem[{{Nolan} {et~al}\mbox{.}(2001){Nolan}, {Dunlop}, {Kukula}, {Hughes},
  {Boroson}, \& {Jimenez}}]{Nolan-2001}
{Nolan} L.~A., {Dunlop} J.~S., {Kukula} M.~J., {Hughes} D.~H., {Boroson} T.,
  {Jimenez} R., 2001, \mnras, 323, 308

\bibitem[{{Ocvirk} {et~al}\mbox{.}(2006){Ocvirk}, {Pichon}, {Lan{\c c}on}, \&
  {Thi{\'e}baut}}]{Ocvirk-2006}
{Ocvirk} P., {Pichon} C., {Lan{\c c}on} A., {Thi{\'e}baut} E., 2006, \mnras,
  365, 46

\bibitem[{{Pacifici} {et~al}\mbox{.}(2013){Pacifici}, {Kassin}, {Weiner},
  {Charlot}, \& {Gardner}}]{Pacifici-2013}
{Pacifici} C., {Kassin} S.~A., {Weiner} B., {Charlot} S., {Gardner} J.~P.,
  2013, \apjl, 762, L15

\bibitem[{Press {et~al}\mbox{.}(1992)Press, Teukolsky, Vetterling, \&
  Flannery}]{NR}
Press W.~H., Teukolsky S.~A., Vetterling W.~T., Flannery B.~P., 1992, Numerical
  recipes in C (2nd ed.): the art of scientific computing. Cambridge University
  Press, New York, NY, USA

\bibitem[{{R Core Team}(2012)}]{R}
{R Core Team}, 2012, R: A Language and Environment for Statistical Computing. R
  Foundation for Statistical Computing, Vienna, Austria, {ISBN} 3-900051-07-0

\bibitem[{{Richards} {et~al}\mbox{.}(2009){Richards}, {Freeman}, {Lee}, \&
  {Schafer}}]{Richards-2009}
{Richards} J.~W., {Freeman} P.~E., {Lee} A.~B., {Schafer} C.~M., 2009, \mnras,
  399, 1044

\bibitem[{{Robertson} {et~al}\mbox{.}(2006){Robertson}, {Hernquist}, {Cox}, {Di
  Matteo}, {Hopkins}, {Martini}, \& {Springel}}]{Robertson-2006}
{Robertson} B., {Hernquist} L., {Cox} T.~J., {Di Matteo} T., {Hopkins} P.~F.,
  {Martini} P., {Springel} V., 2006, \apj, 641, 90

\bibitem[{{S{\'a}nchez-Bl{\'a}zquez}
  {et~al}\mbox{.}(2006){S{\'a}nchez-Bl{\'a}zquez}, {Peletier},
  {Jim{\'e}nez-Vicente}, {Cardiel}, {Cenarro}, {Falc{\'o}n-Barroso}, {Gorgas},
  {Selam}, \& {Vazdekis}}]{Sanchez-Blazquez-2006}
{S{\'a}nchez-Bl{\'a}zquez} P. {et~al.}, 2006, \mnras, 371, 703

\bibitem[{{Schawinski} {et~al}\mbox{.}(2009){Schawinski}, {Virani}, {Simmons},
  {Urry}, {Treister}, {Kaviraj}, \& {Kushkuley}}]{Schawinski-2009}
{Schawinski} K., {Virani} S., {Simmons} B., {Urry} C.~M., {Treister} E.,
  {Kaviraj} S., {Kushkuley} B., 2009, \apjl, 692, L19

\bibitem[{{Sheinis}(2002)}]{Sheinis-2002}
{Sheinis} A.~I., 2002, PhD thesis, University of California, Santa Cruz

\bibitem[{{Smith}(2014)}]{Smith-2014}
{Smith} R.~J., 2014, \mnras, 443, L69

\bibitem[{{Storchi-Bergmann} {et~al}\mbox{.}(2005){Storchi-Bergmann}, {Nemmen},
  {Spinelli}, {Eracleous}, {Wilson}, {Filippenko}, \&
  {Livio}}]{Storchi-Bergmann-2005}
{Storchi-Bergmann} T., {Nemmen} R.~S., {Spinelli} P.~F., {Eracleous} M.,
  {Wilson} A.~S., {Filippenko} A.~V., {Livio} M., 2005, \apjl, 624, L13

\bibitem[{{Thomas} {et~al}\mbox{.}(2005){Thomas}, {Maraston}, {Bender}, \&
  {Mendes de Oliveira}}]{Thomas-2005}
{Thomas} D., {Maraston} C., {Bender} R., {Mendes de Oliveira} C., 2005, \apj,
  621, 673

\bibitem[{{Thomas}, {Maraston} \& {Korn}(2004){Thomas}, {Maraston}, \&
  {Korn}}]{Thomas-2004}
{Thomas} D., {Maraston} C., {Korn} A., 2004, \mnras, 351, L19

\bibitem[{{Tojeiro} {et~al}\mbox{.}(2007){Tojeiro}, {Heavens}, {Jimenez}, \&
  {Panter}}]{Tojeiro-2007}
{Tojeiro} R., {Heavens} A.~F., {Jimenez} R., {Panter} B., 2007, \mnras, 381,
  1252

\bibitem[{{Tokovinin}(2002)}]{Tokovinin-2002}
{Tokovinin} A., 2002, \pasp, 114, 1156

\bibitem[{{Trager} {et~al}\mbox{.}(1998){Trager}, {Worthey}, {Faber},
  {Burstein}, \& {Gonzalez}}]{Trager-1998}
{Trager} S.~C., {Worthey} G., {Faber} S.~M., {Burstein} D., {Gonzalez} J.~J.,
  1998, \apjs, 116, 1

\bibitem[{{Tremonti} {et~al}\mbox{.}(2004){Tremonti}, {Heckman}, {Kauffmann},
  {Brinchmann}, {Charlot}, {White}, {Seibert}, {Peng}, {Schlegel}, {Uomoto},
  {Fukugita}, \& {Brinkmann}}]{Tremonti-2004}
{Tremonti} C.~A. {et~al.}, 2004, \apj, 613, 898

\bibitem[{{Trujillo} {et~al}\mbox{.}(2001){Trujillo}, {Aguerri}, {Cepa}, \&
  {Guti{\'e}rrez}}]{Trujillo-2001}
{Trujillo} I., {Aguerri} J.~A.~L., {Cepa} J., {Guti{\'e}rrez} C.~M., 2001,
  \mnras, 328, 977

\bibitem[{{Trump} {et~al}\mbox{.}(2013){Trump}, {Hsu}, {Fang}, {Faber}, {Koo},
  \& {Kocevski}}]{Trump-2013}
{Trump} J.~R., {Hsu} A.~D., {Fang} J.~J., {Faber} S.~M., {Koo} D.~C.,
  {Kocevski} D.~D., 2013, \apj, 763, 133

\bibitem[{{Vanden Berk} {et~al}\mbox{.}(2001){Vanden Berk}, {Richards},
  {Bauer}, {Strauss}, {Schneider}, {Heckman}, {York}, {Hall}, {Fan}, {Knapp},
  {Anderson}, {Annis}, {Bahcall}, {Bernardi}, {Briggs}, {Brinkmann}, {Brunner},
  {Burles}, {Carey}, {Castander}, {Connolly}, {Crocker}, {Csabai}, {Doi},
  {Finkbeiner}, {Friedman}, {Frieman}, {Fukugita}, {Gunn}, {Hennessy},
  {Ivezi{\'c}}, {Kent}, {Kunszt}, {Lamb}, {Leger}, {Long}, {Loveday}, {Lupton},
  {Meiksin}, {Merelli}, {Munn}, {Newberg}, {Newcomb}, {Nichol}, {Owen}, {Pier},
  {Pope}, {Rockosi}, {Schlegel}, {Siegmund}, {Smee}, {Snir}, {Stoughton},
  {Stubbs}, {SubbaRao}, {Szalay}, {Szokoly}, {Tremonti}, {Uomoto}, {Waddell},
  {Yanny}, \& {Zheng}}]{VandenBerk-2001}
{Vanden Berk} D.~E. {et~al.}, 2001, \aj, 122, 549

\bibitem[{{Walcher} {et~al}\mbox{.}(2006){Walcher}, {B{\"o}ker}, {Charlot},
  {Ho}, {Rix}, {Rossa}, {Shields}, \& {van der Marel}}]{Walcher-2006}
{Walcher} C.~J., {B{\"o}ker} T., {Charlot} S., {Ho} L.~C., {Rix} H.-W., {Rossa}
  J., {Shields} J.~C., {van der Marel} R.~P., 2006, \apj, 649, 692

\bibitem[{{Walcher} {et~al}\mbox{.}(2011){Walcher}, {Groves}, {Budav{\'a}ri},
  \& {Dale}}]{Walcher-2011}
{Walcher} J., {Groves} B., {Budav{\'a}ri} T., {Dale} D., 2011, \apss, 331, 1

\bibitem[{{Wold} {et~al}\mbox{.}(2010){Wold}, {Sheinis}, {Wolf}, \&
  {Hooper}}]{Wold-2010}
{Wold} I., {Sheinis} A.~I., {Wolf} M.~J., {Hooper} E.~J., 2010, \mnras, 408,
  713

\bibitem[{{Wolfram Research, Inc.}(2012)}]{Mathematica}
{Wolfram Research, Inc.}, 2012, {Mathematica}, version 9.0 edn. Wolfram
  Research, Inc., Champaign, Illinois

\bibitem[{{Worthey} {et~al}\mbox{.}(1994){Worthey}, {Faber}, {Gonzalez}, \&
  {Burstein}}]{Worthey-1994}
{Worthey} G., {Faber} S.~M., {Gonzalez} J.~J., {Burstein} D., 1994, \apjs, 94,
  687

\bibitem[{{York} {et~al}\mbox{.}(2000){York}, {Adelman}, {Anderson},
  {Anderson}, {Annis}, {Bahcall}, {Bakken}, {Barkhouser}, {Bastian}, {Berman},
  {Boroski}, {Bracker}, {Briegel}, {Briggs}, {Brinkmann}, {Brunner}, {Burles},
  {Carey}, {Carr}, {Castander}, {Chen}, {Colestock}, {Connolly}, {Crocker},
  {Csabai}, {Czarapata}, {Davis}, {Doi}, {Dombeck}, {Eisenstein}, {Ellman},
  {Elms}, {Evans}, {Fan}, {Federwitz}, {Fiscelli}, {Friedman}, {Frieman},
  {Fukugita}, {Gillespie}, {Gunn}, {Gurbani}, {de Haas}, {Haldeman}, {Harris},
  {Hayes}, {Heckman}, {Hennessy}, {Hindsley}, {Holm}, {Holmgren}, {Huang},
  {Hull}, {Husby}, {Ichikawa}, {Ichikawa}, {Ivezi{\'c}}, {Kent}, {Kim},
  {Kinney}, {Klaene}, {Kleinman}, {Kleinman}, {Knapp}, {Korienek}, {Kron},
  {Kunszt}, {Lamb}, {Lee}, {Leger}, {Limmongkol}, {Lindenmeyer}, {Long},
  {Loomis}, {Loveday}, {Lucinio}, {Lupton}, {MacKinnon}, {Mannery}, {Mantsch},
  {Margon}, {McGehee}, {McKay}, {Meiksin}, {Merelli}, {Monet}, {Munn},
  {Narayanan}, {Nash}, {Neilsen}, {Neswold}, {Newberg}, {Nichol}, {Nicinski},
  {Nonino}, {Okada}, {Okamura}, {Ostriker}, {Owen}, {Pauls}, {Peoples},
  {Peterson}, {Petravick}, {Pier}, {Pope}, {Pordes}, {Prosapio},
  {Rechenmacher}, {Quinn}, {Richards}, {Richmond}, {Rivetta}, {Rockosi},
  {Ruthmansdorfer}, {Sandford}, {Schlegel}, {Schneider}, {Sekiguchi}, {Sergey},
  {Shimasaku}, {Siegmund}, {Smee}, {Smith}, {Snedden}, {Stone}, {Stoughton},
  {Strauss}, {Stubbs}, {SubbaRao}, {Szalay}, {Szapudi}, {Szokoly}, {Thakar},
  {Tremonti}, {Tucker}, {Uomoto}, {Vanden Berk}, {Vogeley}, {Waddell}, {Wang},
  {Watanabe}, {Weinberg}, {Yanny}, {Yasuda}, \& {SDSS
  Collaboration}}]{York-2000}
{York} D.~G. {et~al.}, 2000, \aj, 120, 1579

\end{thebibliography}

\clearpage

\onecolumn
\appendix

\section{Analytic Calculation of Scattered QSO Light}
In this paper, we estimate the scattered quasar light in our host galaxy observations by using an observation of the QSO of that host galaxy, details of the spectroscopic setup, and an estimation of the seeing. With this information, it is possible to estimate the amount of scattered quasar light in the host galaxy observations analytically. The amount of light from a point source into an offset longslit can be readily calculated given the PSF and the specifics of the longslit observation\footnote{Note this can be done in general for any observation geometry, e.g., fibers on an IFU. We focus on longslit as it is the data we deal with in this paper. But, we have developed a similar method for fibers.}. If one assumes that atmospheric turbulence, seeing, dominates the PSF, we can neglect instrumental effects and use a PSF fully described by the seeing. In this paper we use the empirical PSF (Gunn, priv. comm.):

\begin{equation}
\label{psf}
f(r,\sigma) = \frac{9}{13} \left[\frac{1}{2 \pi \sigma^{2}}\exp\left(\frac{-r^{2}}{2\sigma^{2}}\right) \right]+   \frac{4}{13}  \left[\frac{1}{8 \pi \sigma^{2}} \exp\left(\frac{-r^{2}}{8\sigma^{2}}\right)\right]
\end{equation}
The PSF shown here is the sum of two weighted Gaussian with the second Gaussian having a FWHM that is twice the width of the first. This approximates the wings of a PSF dominated by atmospheric seeing. Using a single Gaussian would underestimate the wings. In Eqn. \ref{psf},  $f$ is the normalized 2-D light profile for a point source. The variable $r$ represents the radial distance from the centre of the point source, and the variable $\sigma$ is proportional to the seeing, which is commonly measured and reported as the full-width-half-max of a Gaussian or another suitable profile.

The seeing for large telescopes is typically calculated as \citep{Tokovinin-2002}:
\begin{equation} \label{seeing}
\epsilon_{0} = \frac{0.98 \lambda}{r_{0}}
\end{equation}
Here $\epsilon_{0}$ is the FWHM of a seeing limited profile, $\lambda$ is the wavelength, and $r_{0}$ is the Fried parameter (i.e. the seeing cell size) measured by a differential image motion monitor (DIMM). The Fried parameter has a dependence on wavelength of $r_{0} \propto \lambda^{6/5}$. This means that the seeing, $\epsilon_{0} \propto \lambda^{-1/5}$. Likewise, in our analytic approximation of a seeing limited PSF, the variable $\sigma$, which is directly proportional to the seeing also has a wavelength dependence: $\sigma \propto \lambda^{-1/5}$. If we assume we have a measurement of seeing at some reference wavelength, e.g., $\lambda = 0.5~\mathrm{\mu m}$, we can express $\sigma$ in Eqn. \ref{psf} in its wavelength dependent form.
\begin{equation}
\sigma = \sigma_{0}\left(\frac{\lambda}{\lambda_{0}}\right)^{-1/5}
\end{equation}
Here $\lambda_0$ is the fiducial wavelength and $\sigma_0$ is the Gaussian width (standard deviation) at that wavelength.  In practice the seeing is characterized by the PSF FWHM. Using \textsc{Mathematica} \citep{Mathematica} to solve for the FWHM of this sum of Gaussians, we find:
\begin{equation}
FWHM_{\mathrm{sum}} \sim 2.48~\sigma
\end{equation}
With this understanding of our seeing limited PSF, we only need the additional information detailing the spectroscopic setup. The pertinent details are (1) the width of the slit, $s$, (2) the position of the point source, $(x_{q},y_{q})$, relative to the centre of the slit, $(x_{0},y_{0})$--called the offset, and (3) the extraction width, $e$, (distance perpendicular to dispersion axis) from which light will be collected. 

To find the fraction of light from a point source collected by an extracted longslit spectrum as a function of wavelength, we simply integrate over the area of the extracted slit.
\begin{figure*}
\includegraphics{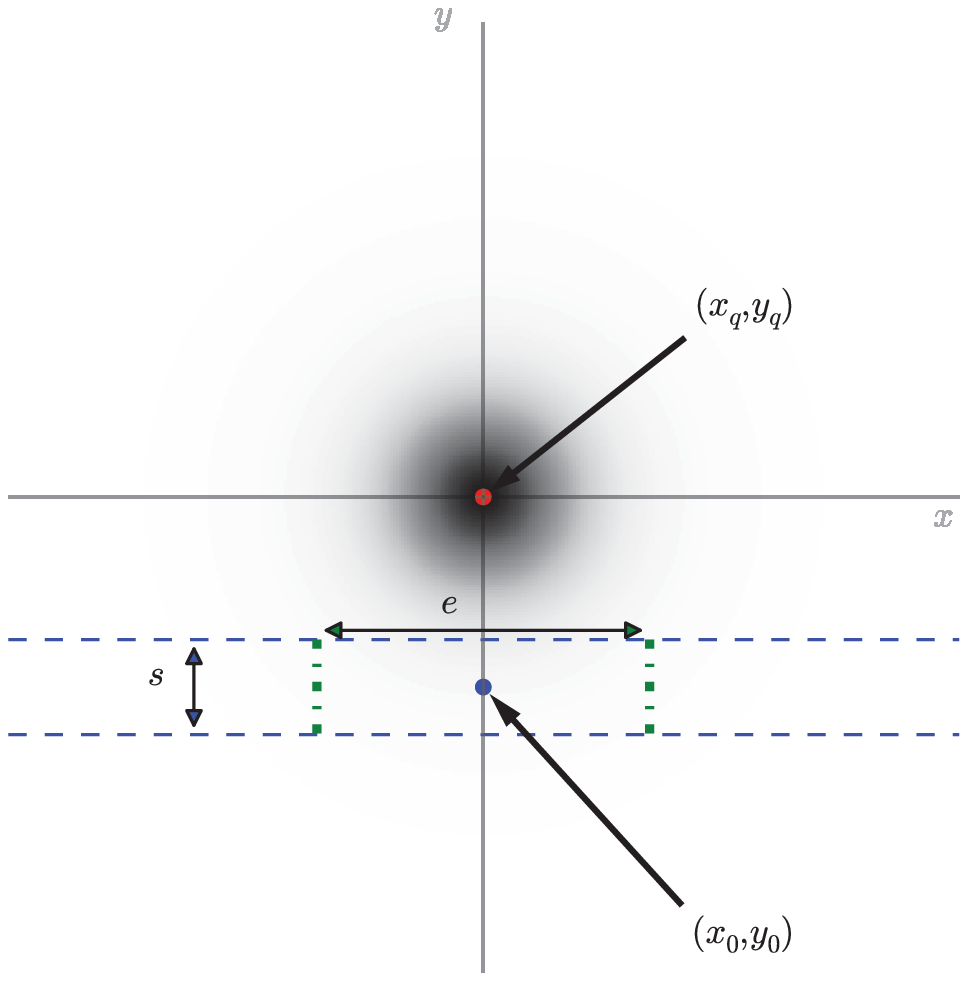}
\caption{Graphical representation of an off-axis longslit observation. The normalized PSF of a point source in 1.7~arcsec seeing is shown in grey-scale. The dashed blue lines represent a 1.0~arcsec slit. The dot-dashed green lines mark the extraction region. The widths and points referenced in the text are displayed.}
\label{fig:slitfig}
\end{figure*}
The particularly simple geometry of longslits (see Fig. \ref{fig:slitfig}) make this an easy integration in Cartesian coordinates:
\begin{equation}\label{light}
L(\sigma,\lambda) = \int_{x_{0}-\frac{e}{2}}^{x_{0}+\frac{e}{2}} \int_{y_{0}-\frac{s}{2}}^{y_{0}+\frac{s}{2}} \frac{9}{13} \left[\frac{1}{2 \pi \sigma^{2}}\exp\left(\frac{-\left[(x-x_{q})^{2}+(y-y_q)^{2}\right]}{2\sigma^{2}}\right) \right]+   \frac{4}{13}  \left[\frac{1}{8 \pi \sigma^{2}} \exp\left(\frac{-\left[(x-x_{q})^{2}+(y-y_q)^{2}\right]}{8\sigma^{2}}\right)\right] dy dx
\end{equation}
This integral can be solved analytically in terms of error functions. Using the analytic solution with error functions is faster than numerical integration, but we found that for even more speed, it is best to make a lookup table that provides the amount of light at a prescribed range of wavelengths for a given value of the seeing.

In this paper, we estimate the scattered quasar flux in the off-axis (host galaxy) observation as the quasar light observed on-axis multiplied by the ratio of the off and on-axis light fractions, $L_{\mathrm{on}}/L_{\mathrm{off}}$ . The setup parameters $s,e,x_q,y_q,x_0,$ and $y_0$ are usually known for both observations, leaving the two seeing values ($\sigma_{\mathrm{on}}$ and $\sigma_{\mathrm{off}}$ ) as the only free parameters. \textsc{sspmodel} steps through a range of seeing values (typically FWHM: 0.3 to 2.0 arcsec) drawing $L_{\mathrm{on}}/L_{\mathrm{off}}$  from a lookup table to estimate the quasar scattered light in the off-axis spectrum. The best fit of the QSO scattered light and the stellar populations to the real data yields the best estimate of contaminating quasar light and scattered light fraction.

\label{lastpage}

\end{document}